%% file: main.tex
\documentclass[submission,copyright]{eptcs}

\providecommand{\event}{MARS 2026}

\usepackage{amsmath}
\usepackage{latexsym}
\usepackage{multirow}

\usepackage{graphicx}
\usepackage{xcolor}
\usepackage[export]{adjustbox}

\definecolor{grey}{cmyk}{0,0,0,0.8}

\usepackage{url}

\hypersetup{bookmarks=true,bookmarksopen=false,colorlinks=true,linkcolor=grey,citecolor=grey,urlcolor=grey}

\usepackage{listings}
\usepackage{cadp-lnt}
\usepackage{cadp-mcl}
\usepackage{cadp-svl}

\newcommand{\B}[1]{\mbox{\bf #1}}
\newcommand{\I}[1]{\mbox{\em #1\/}}

\newcommand{\T}[1]{\mbox{\tt #1}}

\newcommand{\ALT}{$\Box$}
\newcommand{\PAR}{\parallel}
\newcommand{\Q}{\phantom{n}}

\newcommand{\IGNORE}[1]{}

\newcommand{\Picture}[2]{\includegraphics[width=#2\textwidth,valign=m]{#1}}

\newcommand{\SPACING}{\vspace{2ex}}

\title{Guidelines for Producing Concise LNT Models,
Illustrated with Formal Models of the Algorand Consensus Protocol}

\def\titlerunning{Guidelines for Producing Concise LNT Models}

\author{
   Hubert Garavel
   \institute{Univ. Grenoble Alpes, {\sc Inria}, {\sc Cnrs}, Grenoble {\sc Inp},
              {\sc Lig}, 38000 Grenoble, France}
   \email{hubert.garavel@inria.fr}
}

\def\authorrunning{H.~Garavel}

\begin{document}

\maketitle


\begin{abstract}
LNT is a modern language for the formal description of concurrent systems.
It generalizes traditional process calculi and overcomes their known
limitations by incorporating features such as an imperative programming
style with direct assignments to variables, symmetric sequential composition,
and explicit loop operators. The present article examines how these features
can be taken advantage of to obtain LNT models as concise and readable as
possible. The study is illustrated with a running example, the consensus
protocol of the Algorand blockchain, a formal model of which was recently
developed at the University of Urbino. It is shown that, using well-chosen
transformations, the number of lines of LNT code can be divided by three,
while improving readability. Also, various properties of the formal model
are expressed and verified using visual checking, equivalence checking, and
model checking.
\end{abstract}



\section{Introduction}
\label{INTRO}

        Concurrent systems are difficult to design and evolve without the
guidance of formal methods and the assistance of verification tools.
LNT\footnote{\url{https://cadp.inria.fr/tutorial/index.html\#lnt}}
is a modern language for the formal description of such systems.
Inspired by LOTOS \cite{ISO-8807} \cite{Bolognesi-Brinksma-87}, Occam
\cite{May-83} \cite{Hoare-91}, and E-LOTOS \cite{ISO-15437}, LNT pursues
three main objectives:

\begin{enumerate}
        \item Achieving a synthesis between, on the one hand, process calculi,
which have been proposed to model concurrent systems, and, on the other hand,
mainstream (imperative or functional) programming languages, which are
routinely used to develop sequential programs;

        \item Being supported by robust software tools, such as TRAIAN
\cite{TRAIAN-3.17}, a compiler front-end (and C-code generator) that
performs involved static analyses to detect design mistakes in LNT models
as early as possible, and LNT2LOTOS \cite{Garavel-Lang-Serwe-17}
\cite{Champelovier-Clerc-Garavel-et-al-10-v7.5}, a translator from LNT
to LOTOS that enables one to explore the state spaces (Labelled Transition
Systems, LTSs for short) of LNT programs and verify them using the
equivalence checkers and model checkers of the CADP toolbox
\cite{Garavel-Lang-Mateescu-Serwe-13}. To our knowledge, no other modelling
language for concurrent systems benefits from such a wide range of static and
dynamic analyses as available for LNT.

        \item Scaling properly to complex models of real systems, which is
achieved, on the language side, by equipping LNT with programming-in-the-large
features (such as modules, time-proven constructs for structured programming,
Ada-like bracketed syntax, etc.) and, on the software side, by making sure
that the LNT compilers can handle involved programs (for instance, the TRAIAN
and LNT2LOTOS tools are themselves written in LNT, totalling 94,000 non-blank
lines of LNT code).
\end{enumerate}

\noindent
        Since LNT is being used in an increasing number of
case-studies\footnote{\url{https://cadp.inria.fr/case-studies}} and
research tools\footnote{\url{https://cadp.inria.fr/software}}, the question
often arises of how to write LNT code concisely. Indeed, the versatility
of LNT makes it possible to write the same fragment of code in many diverse
ways. The present paper addresses this question, reviewing the different
options, and establishing comparisons with traditional process calculi.

        Throughout the present article, we use Algorand as a running example.
Algorand is a high-performance Layer-1 blockchain that brings convincing
proposals in terms of decentralization, scalability, security, and low-energy
consumption. It is promoted by a foundation\footnote{\url{https://algorand.co}}
and a company\footnote{\url{https://algorandtechnologies.com}}, and supports
two cryptocurrencies (ALGO and EURD). At the heart of Algorand is a secure
distributed ledger algorithm proposed by Chen \& Micali \cite{Chen-Micali-19},
a crucial part of which is the BBA* (generalized Binary Byzantine Agreement)
consensus protocol. A detailed, user-friendly presentation of Algorand can be
found in \cite{Esposito-Rossi-Bernardo-Fabris-Garavel-25}, which also proposes
a formal model of BBA* expressed, first, using process-algebraic notations
and, then, using LNT. We build upon that latter model, the goal being to make
it simpler and more concise through a series of incremental transformations.

The present article is organized as follows.
Section~\ref{MODELLING} introduces the successive LNT models of Algorand
consensus protocol.
Sections~\ref{MODULES}, \ref{TRADEOFFS}, \ref{MERGES}, and \ref{LOOPS} present
four ``generic'' techniques for reducing the size of LNT models.
Section~\ref{VERIFICATION} indicates how these models can be formally
verified using state-space exploration techniques.
Section~\ref{CONCLUSION} gives concluding remarks and suggests directions
for future work.


\section{Formal Modelling of Algorand Consensus Protocol}
\label{MODELLING}

        The starting point for formal models in LNT is the original
description of Algorand~\cite{Chen-Micali-19}, which (unlike many other
blockchains) was not merely a white paper, but a published article in an
established scientific journal. Yet, even if this article was precisely
written and peer reviewed, the description given of Algorand is informal,
contains ambiguities (see Annex~\ref{ANNEX-AMBIGUITIES}), and is not
machine-checkable.

        The next step was the pioneering work done at the University of Urbino
to convert the informal description given by Chen \& Micali into a formal,
process-algebraic model \cite{Esposito-Rossi-Bernardo-Fabris-Garavel-25}
with various simplifying abstractions (see Annex~\ref{ANNEX-ABSTRACTIONS}).
An LNT version of this formal model was produced, which, after a few
iterations\footnote{This explains the successive versions v1--v5 of the arXiv
report \cite{Esposito-Rossi-Bernardo-Fabris-Garavel-25}. The present article
relies on the final version v5.}, passes all the static checks of the LNT
compiler and satisfies basic properties, such as the absence of deadlocks.


\subsection{Specific Transformations}

        Starting from the formal model in LNT given in
\cite{Esposito-Rossi-Bernardo-Fabris-Garavel-25}, we derived a new model
(named version U0) by devising a number of syntactic and semantic
transformations, among which:

\begin{itemize}
        \item improvement of spacing, indentation, and comments;
        \item renaming of identifiers to get more concise or more intuitive names;
        \item introduction of dedicated types for node identifiers and
probability values;
        \item introduction of various functions corresponding to the numeric
constants of \cite{Chen-Micali-19};
        \item tagging of events with observable information (node identifiers,
protocol steps, bit values, etc.);
        \item removal of some events meant for observation, but later found
not so suitable for verification;
        \item displacement at better locations of some events meant for
observation;
        \item merging of two different, yet similar events into a single event;
        \item splitting of a given event into two events to avoid nondeterminism;
        \item adoption of a different attacker model that tries to reject
blocks carrying the bit one.
\end{itemize}

\noindent
        These transformations, detailed in Annex~\ref{ANNEX-SPECIFIC}, are
called ``specific'' because they largely depend on the formal model of
Algorand and are not directly reusable for other applications. Many of these
transformations modify the transitions or transition labels of the LTSs
generated from the LNT code, and thus do not preserve bisimilarity properties.


\subsection{Generic Transformations}
\label{GENERIC}

        Reducing the size (measured in lines of code) of formal models is
a suitable goal. It is expected to reduce the effort spent in reading and
understanding these models, correcting their mistakes, maintaining them
over time, and implementing them to produce real systems.

        In this section and the next ones, we focus on transformations that
we call ``generic'' as they do not modify the semantics of LNT models.
Precisely, such transformations should preserve strong bisimilarity between
the models before and after transformation.

        A simple generic transformation to reduce the size of formal models
consists in removing useless definitions, i.e., all objects defined but not
used (this is likely to occur when a model undergoes many successive
evolutions). The TRAIAN compiler for LNT helps to detect such cases of
over-specification by warning about all types, variables, functions, processes,
etc. that are not actually useful.

        To reduce the size of formal models, other generic transformations
exist. Four of them are presented in the next Sections~\ref{MODULES}
to~\ref{LOOPS} and applied to version U0 of the Algorand consensus protocol,
leading to four successive versions noted U1, U2, U3, and U4. Thus, the
five LTSs generated for versions U0 to U4 are expected to be strongly
bisimilar. Using the BCG\_CMP
tool\footnote{\url{https://cadp.inria.fr/man/bcg_cmp.html}} of CADP, we
verified this property on the two Algorand configurations $A_{4,0}$ and
$A_{2,2}$ studied in \cite{Esposito-Rossi-Bernardo-Fabris-Garavel-25},
where $A_{H,M}$ denotes a network with $H$ honest nodes and $M$ malicious
nodes.

        The table below gives comparative information about the five versions
U0 to U4, both in lines of code for the various LNT models and in numbers
of states for the corresponding LTSs (generated using CADP). In this table,
$L$ is the total number of LNT lines; $\ell$ is the total number of LNT
lines, excluding blank lines and comments; $S_{1,0}$ is the number of states
of the LTS generated  for one honest node; $S_{0,1}$ is the number
of states of the LTS generated for one malicious node; $S_{4,0}$ is the number
of states of the LTS generated for $A_{4,0}$; and $S_{2,2}$ is the number of
states of the LTS generated for $A_{2,2}$. The last line of the table gives
the numbers of states for these four LTSs minimized wrt strong bisimulation,
these numbers being identical for all five versions.

\begin{center}
\begin{tabular}{|c||c|c||c|c||c|c|} \hline
version & $L$ & $\ell$ & $S_{1,0}$ & $S_{0,1}$ & $S_{4,0}$ & $S_{2,2}$ \\ \hline \hline
U0 & 769 & 705 & 1740 & 3190 & 24,230 & 42,509 \\ \hline
U1 & 607 & 552 & 1740 & 3190 & 24,230 & 42,509 \\ \hline
U2 & 320 & 282 & 1761 & 3190 & 24,599 & 42,509 \\ \hline
U3 & 288 & 250 & 1681 & 3046 & 20,665 & 34,679 \\ \hline
U4 & 250 & 212 & 1723 & 3130 & 20,905 & 35,059 \\ \hline \hline
\multicolumn{3}{|c||}{strongly minimized LTSs:} & 558 & 840 & 12,059 & 19,486 \\ \hline
\end{tabular}
\end{center}


\section{Model Compaction Through Modules}
\label{MODULES}

        Quite often, formal models contain identical code fragments that the
specifier replicated using cut-and-paste facilities. Also, various instances
of a formal model may contain identical code fragments located in different
computer files. Such a duplication of code is a true nuisance: it is easy to
create, but has enormous costs on the long run in terms of bug fixes and
maintenance.

        Modularity is a proper way to fight code duplication. Pioneering
programming languages (e.g., Ada and Standard ML) offer sophisticated modules
with interfaces and generics. Similar ideas
\cite{Brinksma-88}
\cite{Brinksma-Leih-95}
\cite{Garavel-Sighireanu-96-e}
have been put forward for the design of E-LOTOS, but never actually
implemented.

        LNT currently proposes a lighter approach to modules, which
addresses most practical needs. Each LNT module is a file
containing definitions of types, functions, processes, and/or
channels\footnote{Channels specify the types of communication events.}.
Modules may import other modules: the import relation may be tree-like or
dag-like, but may not contain cycles. Each module is self-contained, meaning
that it may only refer to objects defined in itself, in the modules it
imports, or in the predefined LNT library. Name clashes are forbidden, but
this constraint is alleviated by the existence of separate name spaces and
the possibility to define overloaded functions. Finally, LNT provides for
generic modules parameterized by ``virtual'' types, functions, processes,
and/or channels.

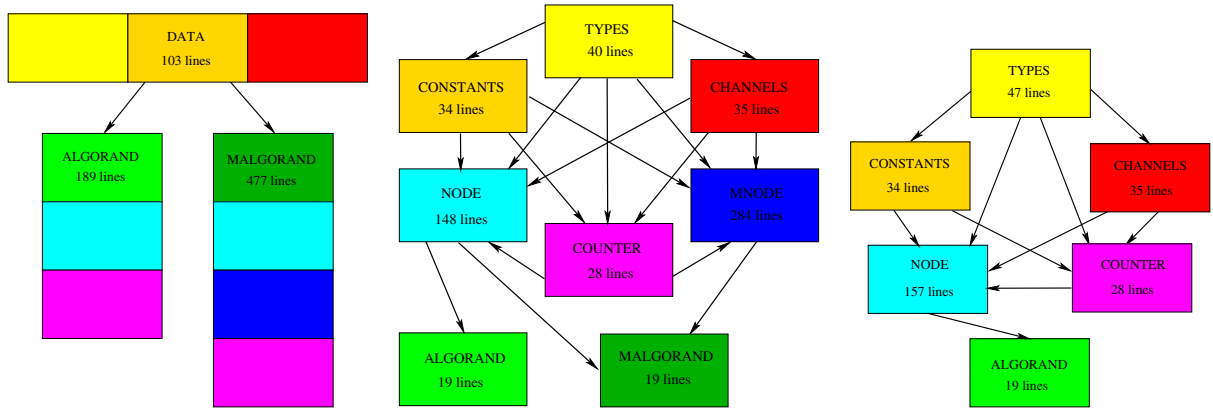
\begin{figure}
\begin{center}
\resizebox{0.30\textwidth}{!}{\input{fig_u0}} \hfill
\resizebox{0.35\textwidth}{!}{\input{fig_u1}} \hfill
\resizebox{0.30\textwidth}{!}{\input{fig_u2}}
\end{center}
\caption{Modular decompositions of versions U0 (left), U1 (middle), and U2 (right)}
\label{FIG-MODULES}
\end{figure}

        Modules were effective in reducing the size of Algorand models.
Version U0 was already decomposed in three modules: \T{ALGORAND.lnt} and
\T{MALGORAND.lnt}, which describe the configurations $A_{4,0}$ and $A_{2,2}$,
respectively, and \T{DATA.lnt}, which defines types and functions used by
these two modules. Yet, as shown in Fig.~\ref{FIG-MODULES} (left), the two
former modules contain two identical fragments of LNT code, namely the
definition of a honest Algorand node (light-blue box) and the definition of
a counter process (purple box). This prompted for the introduction of two
modules that avoid such duplication of code. Additionally, module \T{DATA.lnt}
was split in three smaller modules (types, constants, and channels) and
a new module \T{MNODE.lnt} was created to contain the definition of a
malicious Algorand node. This led to version U1, shown in
Fig.~\ref{FIG-MODULES} (middle), where \T{ALGORAND.lnt} and \T{MALGORAND.lnt}
remain as two tiny modules (19~lines each) that express the top-level
architecture. When evolving from version U0 to version U1, the number of
lines of LNT code was reduced by more than 20\%, while the sizes of LTSs
remained unchanged (see Table in Sect.~\ref{GENERIC}).

\section{Model Compaction Through Control/Data Tradeoffs}
\label{TRADEOFFS}

        Sequential programs often contain multiple fragments of code that,
although not, strictly speaking, identical, are largely similar, differing
only in a few details. To reduce the size of such programs and make them
easier to maintain, the traditional approach is ``procedural abstraction'',
which encapsulates replicated fragments of code within reusable procedures
or functions equipped with appropriate parameters that take care of the few
differences between code fragments.

        The same approach applies to formal models of concurrent systems,
with the additional fact that executing one particular code fragment is an
information that plays a role in the global state of the system (as analyzed
by explicit-state and symbolic verification). Indeed, the global state
usually contains both control information (e.g., location of the program
counter, current states in automata, marked places of a Petri net, etc.)
and data information (e.g., current value of state variables). In this
respect, procedural abstraction not only factors out replicated code
fragments, but also converts control information to data information in
the global state.
        For instance, if $B_0$ and $B_1$ are two similar LNT code fragments,
the nondeterministic choice ``\B{alt} $B_0$ \ALT\ $B_1$ \B{end alt}''
may equivalently be written ``$i$ := \B{any} \I{BIT}\:; $B_i$'' using 
nondeterministic selection of a bit value, thus converting control
information (program counter located in either $B_0$ or $B_1$) into data
information (the value of bit $i$\/).

        Replicated code fragments can be found in version U1 of the Algorand
model: the \T{NODE.lnt} module contains six processes (\I{NODE}, $N$, $N'$,
$N''$, $N'''$, and $N''''$) describing a honest node, whereas the
\T{MNODE.lnt} module contains six processes (\I{MNODE}, $M\!N$, $M\!N'$,
$M\!N''$, $M\!N'''$, and $M\!N''''$) describing the offensive behaviour of
a malicious node and four processes ($H\!N'$, $H\!N''$, $H\!N'''$, and
$H\!N''''$) describing the neutral behaviour of a malicious node. Procedural
abstraction removes the need for \T{MNODE.lnt} by merging these processes as
follows: \{\I{NODE}, \I{MNODE}\}, \{$N$, $M\!N$\},
\{$N'$, $M\!N'$, $H\!N'$\}, \{$N''$, $M\!N''$, $H\!N''$\},
\{$N'''$, $M\!N'''$, $H\!N'''$\}, and \{$N''''$, $M\!N''''$, $H\!N''''$\}
after giving them an additional parameter $M$ that may take three values
(\I{honest}, \I{malicious}, or \I{disguised}) and, when equal to \I{malicious},
forces nodes to cast rejection votes.
        Rather than merging and modifying these processes all at once, with
a high risk of errors, the changes were done incrementally, by applying a
dozen simple transformations in sequence and checking that strong
bisimilarity was preserved at each step.

        The model can be further simplified by noticing that, even if each
node is assigned a distinct number \I{ID}, this number has no real impact on
the behaviour of honest and malicious nodes (this is related to the ideas
of data independence and symmetry reduction of \cite{Ip-Dill-96}). Without
loss of generality, one can assign the lowest numbers to honest nodes, so
that comparing the \I{ID} parameter of each node against the number \I{H} of
honest nodes determines if the node is honest or not. Thus, module
\T{MALGORAND.lnt} is no longer useful and can be eliminated, leading to the
version~U2 shown in Fig.~\ref{FIG-MODULES} (right). With respect to version
U1, the number of lines of LNT code in version U2 was divided by two
(see Table in Sect.~\ref{GENERIC}).


\section{Model Compaction Through Common-Sequence Merges}
\label{MERGES}

        Another cause of verbosity in formal models of concurrent systems
lies in the particular history of process calculi. Initially, the pioneering
CSP language defined by Hoare \cite{Hoare-78} in 1978 had the usual (i.e.,
symmetric) sequential composition operator (noted ``;'') that already existed
in most imperative programming languages. A major shift occurred two years
later when Milner proposed its CCS calculus \cite{Milner-80}, in which
sequential composition had to be expressed with an asymmetric ``action
prefix'' operator. Theoretically, CCS offered some advantages: a tiny syntax,
a concise semantics, and a sufficient expressiveness. But it had many
practical drawbacks: (i) it did not scale well to large systems --- actually,
few significant case-studies have been done using CCS; (ii) its sequential and
parallel operators fragmented the scientific community into separate schools
of thought, e.g., those who adopted action prefix, those who rejected it,
and those who tried to have both action prefix and symmetric sequential
composition in the same language; (iii) it created a gap between process
calculi and conventional programming languages, durably isolating concurrency
theory from practical applications and letting lower-level formalisms, such
as state machines, take over process calculi.

        With respect to model compactness, the action prefix operator of CCS
naturally leads to duplicated code fragments, as sequential processes must
have a tree-like structure (based on action prefix and nondeterministic choice)
with no possibility of dag-like factorization. For instance, ``$(b_1+b_2).B$''
is forbidden in CCS and must be written ``$(b_1.B)+(b_2.B)$\/'' instead.

        The LNT language has a value-passing sequential composition operator
that can express the action prefix of CCS as a particular case
\cite[Sect.~5.3]{Garavel-15-b}. When models written in the CCS style are
directly translated to LNT, duplicated code fragments are likely to be
present, e.g., ``\B{alt} $b_1$; $B$ \ALT\, $b_2$; $B$ \B{end alt}''
(same for ``\B{if} ... \B{end if}'' and ``\B{case} ... \B{end case}''
statements). Fortunately, the sequential composition operator of LNT allows
duplicated code fragments to be factored out, which was not permitted in CCS.

        Version U2 of the Algorand model uses the CCS style, since this LNT
program was derived from the process-algebraic specification presented in
\cite{Esposito-Rossi-Bernardo-Fabris-Garavel-25}. Hence, there are
duplicated sequences of code and recursive process calls in version U2.
To reduce their number, each of the five processes $N$, $N'$, ..., $N''''$
of \T{NODE.lnt} was modified separately, using the following transformations
(the corresponding code fragments of versions U2 and U3 can be found in
Annexes~\ref{ANNEX-U2-NODE} and~\ref{ANNEX-U3-NODE}, respectively):

\begin{itemize}
        \item For process $N$: the transformation is trivial. It introduces
an auxiliary variable $M$ and replaces the three calls to process $N'$ by a
single call.

        \item For process $N'$: the transformation is also trivial. It
introduces an auxiliary variable $B$ and replaces the two calls to process
$N''$ by a single call

        \item For process $N''$: the transformation requires two successive
steps. First, each of the three ``\B{alt}'' statements is simplified by
applying the following law:\footnote{Notice that this law only holds for
common suffixes and would not hold for common prefixes.}
\begin{center}
\B{alt} $B_1$; $B$ \ALT\, $B_2$; $B$ \B{end alt} ~$=$~
\B{alt} $B_1$ \ALT\, $B_2$ \B{end alt}\,; $B$
\end{center}
where ``$=$'' denotes strong bisimilarity and where the common suffix $B$
corresponds to the code fragments starting at ``SYNC (END)''.
Then, the three branches of the modified ``\B{case}'' statement are
simplified by applying the following law:
\begin{center}
\B{case} 0 $\rightarrow$ $B$; $B_0$ $\mid$ 1 $\rightarrow$ $B$; $B_1$ $\mid$
2 $\rightarrow$ $B$; $B_2$ \B{end case} ~$=$~ $B$; \B{case} 0 $\rightarrow$
$B_0$ $\mid$ 1 $\rightarrow$ $B_1$ $\mid$ 2 $\rightarrow$ $B_2$ \B{end case}
\end{center}
where the common prefix $B$ corresponds to the code fragments starting at
``\B{alt}'' and ending with ``\B{end alt}; SYNC (END)''. After these
transformations, process $N''$ is still called twice in version U2; it could
be factored out, but the resulting code would be longer and less readable.

        \item For process $N'''$: the transformation exploits the fact that
the two executable branches of the ``\B{case}'' start with a common prefix
(the TALLY event) followed by two largely similar code fragments.

        \item For process $N''''$: the transformation exploits the fact that
the three branches of the ``\B{case}'' bear similarities: the two first
branches are dual of each other, and the third branch borrows code from
the two other branches. The transformation introduces two auxiliary
variables \I{DONE} and $B$ and replaces the six calls to process $N''$ by a
single call. The resulting code is concise, yet harder to understand (its
correctness can be checked by executing symbolically both versions of process
$N''''$ three times, one per value of $S$). The transformation can be seen as
an exercise to get a different view at the code of process $N''''$, possibly
suggesting further refactoring of the whole node process.
\end{itemize}

\noindent
Again, these transformations were done one at a time, carefully
checking the preservation of strong bisimilarity after each change.
Between versions U2 and U3, the number of process calls in \T{NODE.lnt} was
reduced from~22 to~9. The number of lines of LNT code was reduced by 10\%,
with a noticeable effect on state spaces: 16\% less states for $S_{4,0}$
and 19\% less states for $S_{2,2}$ (see Table in Sect.~\ref{GENERIC}).


\section{Model Compaction Through Explicit Loops}
\label{LOOPS}

        Our last transformation addresses two causes of superfluous verbosity
in models of concurrent systems:

\begin{itemize}
        \item Communication protocols are often described using state machines,
which are defined using control states, transitions between these states,
and state variables, whose values are checked and modified by code fragments
written in imperative programming languages (as, e.g., in Estelle
\cite{ISO-9074} or SDL \cite{ITU-Z100-99}). Unfortunately, we know from
Dijkstra that gotos can be harmful \cite{Dijkstra-68-b} and, indeed, state
machines may easily lead to ``spaghetti code'', in which the control flow
is difficult to follow, especially because loops are not identified explicitly.

        \item The programming style proposed by Milner for CCS (and later
adopted by process calculi such as ACP \cite{Bergstra-Klop-84} and LOTOS
\cite{ISO-8807}) radically differs from that of CSP in two more points:
(i)~CCS has no loop operator, so that any iteration must be expressed using
process recursion; (ii)~CCS is strictly functional, without explicit
assignments to variables. Consequently, in the CCS style, protocols can only
be specified using state machines, in a verbose and non-intuitive manner.
Precisely, each state machine with $n$ states is encoded by $n$ processes,
and each transition from state $s_1$ to state $s_2$ is encoded by a
tail-recursive call of the process corresponding to $s_2$. Moreover, each
state variable of this machine is declared $n$ times (as a parameter of each
of these processes), and $n^2$~parameter passings are needed to encode the
assignments to state variables, even in the cases where these variables are
not modified\footnote{The formal model given in
\cite[Sect.~3.3]{Esposito-Rossi-Bernardo-Fabris-Garavel-25} manages to
conceal verbosity by writing these parameters as tiny subscripts of processes
$N$, $N'$, ..., $N''''$ and by not declaring their types, letting readers
infer unspecified types.}.
\end{itemize}

\noindent
        To improve the quality and compactness of formal models, it is thus
desirable to migrate away from the CCS-like style, replacing parameter passing
by direct assignments to state variables, and replacing gotos (encoded as
tail-recursive process calls) by the higher-level control structures
(loops, ``\B{if-then-else}'' conditionals, etc.) inherited from structured
programming.

        Fortunately, the LNT language is versatile enough to support both the
CCS-like functional/recursive style and the imperative style with loops and
variable assignments. The main reason for the functional style was to ensure
that each variable is properly assigned before used (a prerequisite for
having a formal semantics), but LNT resolves this issue in a much more
flexible way, by using static analysis \cite{Garavel-15-b}.

        In version U3 of the Algorand model, two modules (\T{COUNTER.lnt} and
\T{NODE.lnt}) are written in the CCS-like functional/recursive style.

        Concerning \T{COUNTER.lnt}, the transformation to imperative style
is straightforward (the corresponding code fragments can be found in
Annexes~\ref{ANNEX-U3-COUNTER} and~\ref{ANNEX-U4-COUNTER}, respectively).
The three recursive calls in version U3 are replaced by one loop and
assignments to variables in version U4. Both versions have nearly the
same number of LNT lines: deciding which one is easier to read is really
a matter of individual taste. A minor advantage of version U4 is the strict
encapsulation of both variables $K_0$ and $K_1$ in the \I{COUNTER} process,
while version U3 exports these variables as parameters that need to be
initialized by the caller process. This well-known issue with traditional
process calculi could be resolved either by defining an additional process
(without parameters) that calls the \I{COUNTER} process to initialize its
$K_0$ and $K_1$ parameters, or by extending the LNT language with default
values for parameters (as, e.g., in Ada or Python), which we prefer to avoid
as it would interfere with overloading.

        Concerning \T{NODE.lnt}, it is clear that the five mutually
recursive processes $N$, $N'$, ..., $N''''$ define one single state machine,
since all process calls contained in them never return and rather express
gotos between states than genuine process calls with a stack-based semantics.
Thus, processes $N$, $N'$, ..., $N''''$ must be considered and transformed
altogether, as they form a unique piece of sequential code.

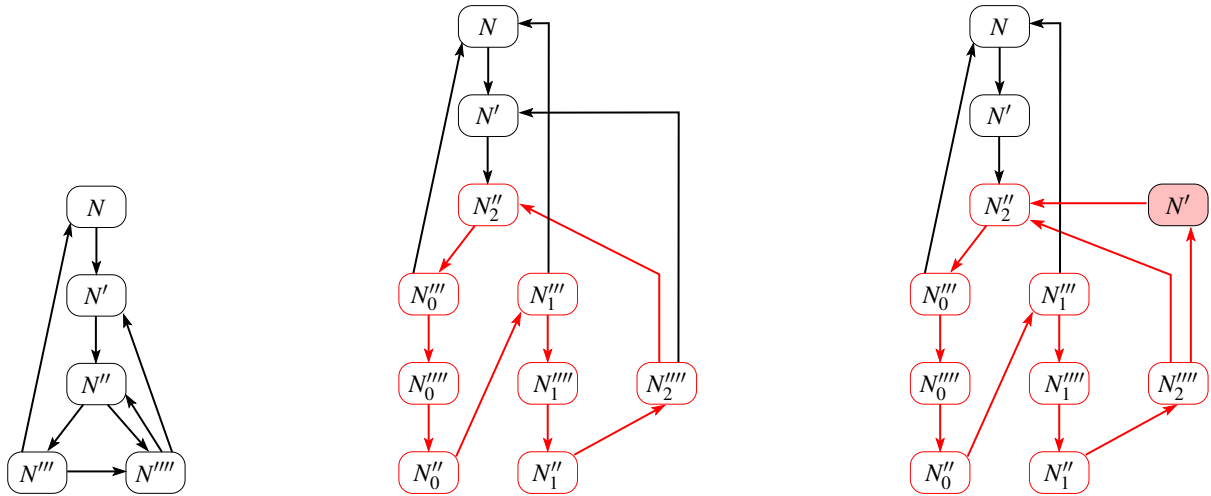
\begin{figure}
\begin{center}
\resizebox{0.15\textwidth}{!}{\input{fig_u4a}} \hfill
\resizebox{0.25\textwidth}{!}{\input{fig_u4b}} \hfill
\resizebox{0.25\textwidth}{!}{\input{fig_u4c}}
\end{center}
\caption{Loop discovery: call graph for processes only (left); call graph for
pairs of processes and steps (middle); call graph for pairs processes and
steps, with duplication of process $N'$ (right)}
\label{FIG-LOOP}
\end{figure}

        At first sight, the call graph of these processes, shown on
Fig.~\ref{FIG-LOOP} (left), is quite involved, so that using LNT loops
instead of gotos is unlikely to increase conciseness and readability.
But, a comparison with the algorithm of \cite{Chen-Micali-19} suggests
that the ``true'' states of the state machine may not merely be the five
control locations $N$, $N'$, ..., $N''''$. Instead, the ``true'' states
seem to be pairs $(L, S)$, where $L$ is one of these five control locations
and $S$ is the current step, either 0, 1, or 2 (the latter value being
also noted \I{S\_INIT}).

        A symbolic exploration of all reachable states $(L, S)$ gives the
call graph shown in Fig.~\ref{FIG-LOOP} (middle), where (for conciseness of
the figure) $L_S$ denotes process $L$ called with step parameter $S$,
while processes $N$ and $N'$, which have no step parameter, are simply noted
$N$ and $N'$. Doing so, the inherent structure of the consensus algorithm
appears, with an inner loop (in red) that cyclically executes the three
``steps'' defined in \cite{Chen-Micali-19} (\I{Coin-Fixed-To-0},
\I{Coin-Fixed-To-1}, and \I{Coin-Genuinely-Flipped}), and an outer loop
(in black) that contains the inner loop and performs an infinite sequence
of ``rounds'', each of which receives and processes a new block.

         Therefore, the consensus algorithm can easily be expressed using
two nested LNT loops, the inner loop having three ``\B{break}'' statements
that exit to the outer loop and correspond, in Fig.~\ref{FIG-LOOP} (middle),
to the three bottom-up edges displayed in black. Two of these edges go back
to $N$, while the third edge goes back to $N'$: this can easily be
expressed in LNT by introducing a Boolean variable $X$ that distinguishes
between $N$ and $N'$ and is assigned before each ``\B{break}'' statement,
together with an ``\B{if}-\B{then}'' conditional that only executes $N$ if
$X$ has a given value, or jumps directly to $N'$ otherwise.

        Yet, one may notice that process $N'$, which computes a random bit
equal to zero with probability $P$, is called twice and plays a double role:
(i) when called from process $N$ with $P=P_H$, it models the result of
Algorand's graded consensus phase (not modelled in detail) that takes place
at the beginning of the outer loop, and (ii) when called from process
$N''''_2$ with $P=0.5$, it expresses the probabilistic choice of a random
bit that takes place during the 3rd step (\I{Coin-Genuinely-Flipped}) of
Algorand's consensus algorithm (see Annex~\ref{ANNEX-U2-NODE} for details).
The factorization of (i) and (ii) in a single process $N'$ was a design
decision of the initial formal model
\cite{Esposito-Rossi-Bernardo-Fabris-Garavel-25} --- which wrongly called
$N'$ with $P=P_H$ in the case (ii). Such a factorization somehow altered
the structure of the algorithm by moving case (ii) before the inner loop,
whereas case (ii) logically belongs to the inner loop, of which it is the
last step.

        Given that process $N'$ is small, we decided to revert this
factorization and have instead two copies of $N'$, as it was originally the
case in \cite{Chen-Micali-19}. The resulting call graph is shown in
Fig.~\ref{FIG-LOOP} (right). The translation to LNT becomes simpler, as
the inner loop now has only two ``\B{break}'' statements, and one no longer
needs the aforementioned Boolean variable $X$ nor the additional
``\B{if}-\B{then}'' conditional.

        Consequently, the module \T{NODE.lnt} was fully rewritten in version
U4 by removing the four processes $N'$, ..., $N'''$ after expanding their
contents inline in process $N$, dispatching the various branches of the
``\B{case}'' statements where needed. The module \T{TYPES.lnt} was also
simplified by removing the type \I{STEP} and two functions \I{NEXT} and
\I{S\_INIT}, which are no longer used in version U4. To factor out
duplicated code, the new process $N$ can then be shortened by introducing
three auxiliary processes that contain neither loops nor process calls:

\begin{itemize}
        \item Process \I{FIX\_COIN} encapsulates assignments to the bit $B$,
noticing that each such assignment is always followed by a \I{SET\_BIT} event.

        \item Process \I{FLIP\_COIN} corresponds to the former (duplicated)
process $N'$ (i.e., the probabilistic computation of a random value for the
bit $B$) at the end of which the initial \I{SET\_BIT} event of process $N''$
was added.

        \item Process \I{BROADCAST} corresponds to the former process $N''$,
from which the initial \I{SET\_BIT} event was removed, and at the end of
which the initial \I{TALLY} event of process $N'''$ was added.
\end{itemize}

\noindent
        Thus, version U4 deeply changed the appearance of the model, making
it look much closer to the algorithm of \cite{Chen-Micali-19} and bringing
new insight into Algorand. Compared to version U3, the 12~recursive process
calls present in the \T{COUNTER.lnt} and \T{NODE.lnt} modules have been
replaced by three LNT loops, and the number of lines of LNT code was further
reduced by 13\% (see Table in Sect.~\ref{GENERIC}).


\section{Formal Verification}
\label{VERIFICATION}

        As mentioned in Sect.~\ref{GENERIC}, the LTSs corresponding to
versions U0, ..., U4 are strongly bisimilar. Can one obtain further
guarantees that these LNT models faithfully describe the Algorand consensus
protocol?

        A part of the answer lies the LNT compiler, which performs multiple
checks based on control-flow and data-flow analyses. In earlier versions
of the Algorand model, these checks detected various problems (e.g., wrong
synchronizations likely to cause deadlocks), which have been fixed in
\cite{Esposito-Rossi-Bernardo-Fabris-Garavel-25} and U0.

        Another part of the answer lies in state-space exploration methods
(i.e., LTS construction) which, although expensive, are easily tractable for
Algorand configurations with 4~nodes (see Table in Sect.~\ref{GENERIC}).
We now briefly present how such configurations can be analyzed using visual
checking, equivalence checking, and model checking. Larger configurations
have also been explored using the compositional verification
\cite{Garavel-Lang-Mounier-18} capabilities of CADP (e.g., 6~nodes in
Grenoble, 8 and 10 nodes in Urbino).


\subsection{Visual Checking}
\label{VISUAL-CHECKING}

        The LTSs generated for Algorand with 4~nodes have, after minimization
for strong bisimulation, 12,000--20,000 states, 43,000--69,000 transitions, and
131 different labels. They are too complex for visual inspection by a human.
One approach is to use ``event slicing'' abstractions, i.e., hiding (or
renaming) all events but a few ones of interest, minimizing the LTS for
branching bisimulation, and observing the minimized LTS, if it is small
enough to be visually inspected. For instance, if only \I{SYNC} events are
kept visible, the minimized LTS is an infinite loop of \I{SYNC (BEGIN)} events
followed by \I{SYNC (END)} events, which gives a positive indication of
correctness.

\begin{figure}
\begin{center}
\resizebox{0.18\textwidth}{!}{\input{fig_vc_d1}} \hfill
\resizebox{0.18\textwidth}{!}{\input{fig_vc_d2}} \hfill
\resizebox{0.2\textwidth}{!}{\input{fig_vc_ok}} \hfill
\resizebox{0.16\textwidth}{!}{\input{fig_vc_ko}}
\end{center}
\caption{Visual checking -- (a) and (b): deadlocks, (c): normal behaviour,
(d): corrupted behaviour}
\label{FIG-VC}
\end{figure}
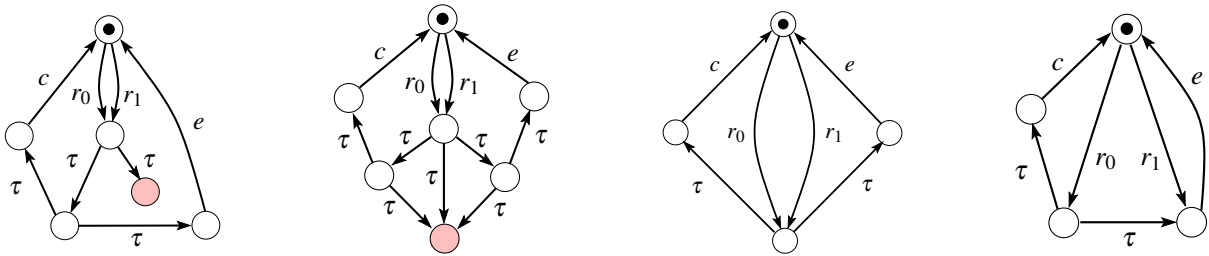

        If only the four events corresponding to the upper-level interface of
the consensus protocol are kept visible, namely,
\I{RECEIVE\_BLOCK\_PROPOSAL (0 or 1)}, \I{COMMIT\_PROPOSED\_BLOCK}, and
\I{COMMIT\_EMPTY\_BLOCK} (which we abbreviate as $r_0$, $r_1$, $c$, and $e$,
respectively), other interesting properties can be discovered using visual
checking. For instance, the presence of deadlocks in earlier versions of
the protocol\footnote{i.e., prior to version v5 of the arXiv report
\cite{Esposito-Rossi-Bernardo-Fabris-Garavel-25}.}
can be seen easily (red states) in Fig.~\ref{FIG-VC}~(a) and~(b) ---
the CADP tools then give the shortest sequences of events leading to these
deadlock states.
        Also, by varying the number of honest nodes in 0...$N$ and the
threshold value $T$ (used by the vote algorithm) in 1...$N$, one observes
that, if $H \geq T$, the minimized LTS is that of Fig.~\ref{FIG-VC}~(c),
which shows a normal behaviour where every block received (event $r_0$
or $r_1$) is either accepted (event $c$) or rejected\footnote{The rejection
of a proposed block in Algorand is expressed by committing an empty block.}
(event $e$). But, if $H < T$, i.e., if there are too many malicious nodes,
the minimized LTS is that of Fig.~\ref{FIG-VC}~(d), in which attacks are
clearly successful, since any block carrying the bit one received is
systematically rejected. More examples of visually checked properties are
given in Annex~\ref{ANNEX-PROPERTIES}.

\subsection{Equivalence Checking}
\label{EQUIVALENCE-CHECKING}

        Equivalence checking \cite{Garavel-Lang-22}, as it is used in the
present article, is a verification technique based on the comparison of two
LTSs to decide whether they are bisimilar (e.g., for strong or branching
bisimulation) or whether one LTS is included in the other for some behavioural
preorder relation. As mentioned above, equivalence checking was already used
to prove that the successive versions of the Algorand model are strongly
bisimilar. But equivalence checking can be used more widely, e.g., to
check whether a complex system (namely, an LTS generated from a formal
model of Algorand, after hiding and/or renaming certain labels to keep only
those events of interest) is branching bisimilar to a simpler system, the
correctness of which is evident.

\begin{figure}
\begin{center}
\begin{small}
\hspace*{-2mm}
\Picture{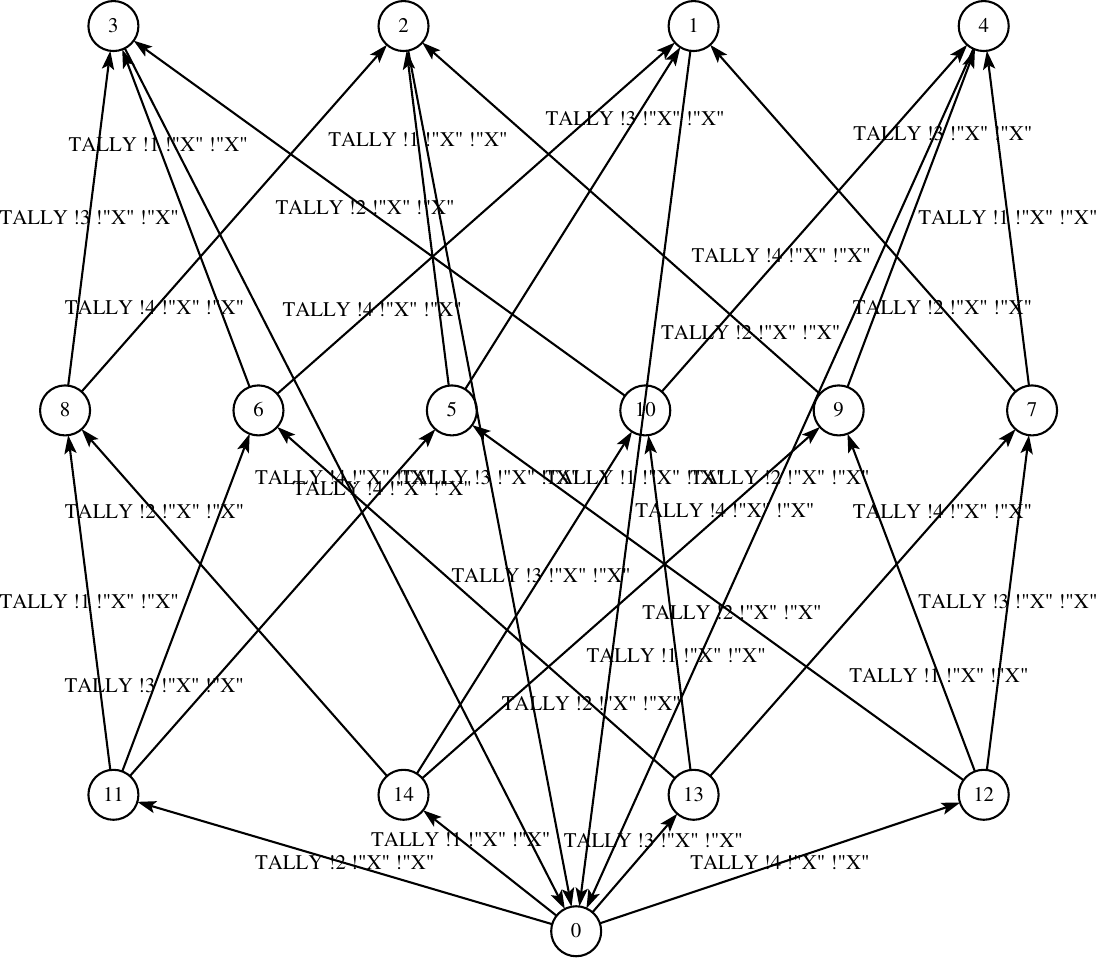}{0.2} 
\begin{tabular}{l}
\B{loop} \\
\Q \B{par} \\
\Q \Q tally $(1, X, X)$ \\
\Q \Q $\PAR ... \PAR$ \\
\Q \Q tally $(4, X, X)$ \\
\Q \B{end par} \\
\B{end loop}
\end{tabular} 
\Picture{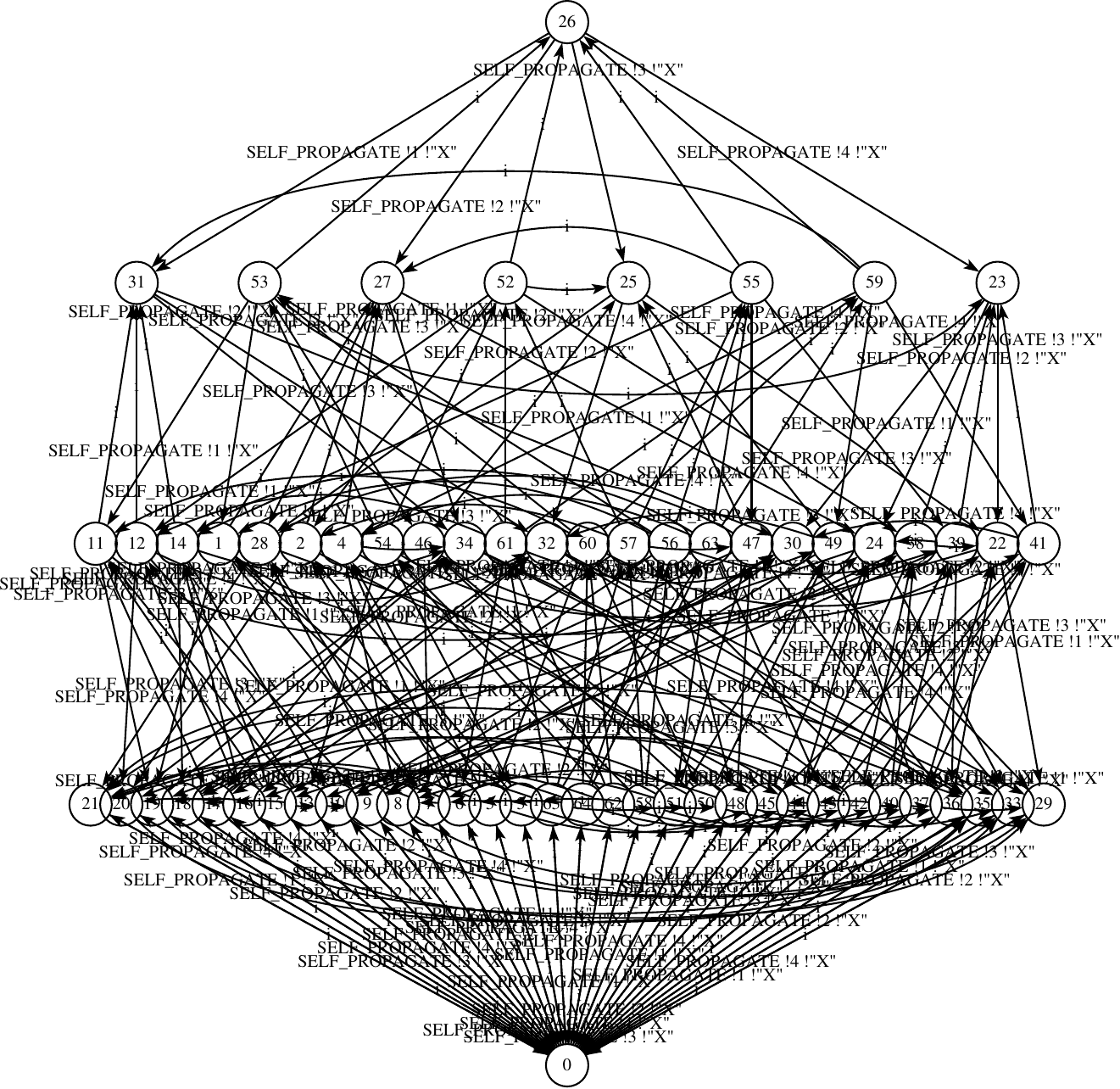}{0.2}
\begin{tabular}{l}
\B{loop} \\
\Q \B{par} \\
\Q \Q $p(1) \PAR ... \PAR p(4)$ \\
\Q \B{end par} \\
\B{end loop} \\
~ \\
~ \I{where} \\
\end{tabular}
\begin{tabular}{l}
\B{process} $p (n)$ \B{is} \\
\Q $\tau$; \\
\Q \B{alt} \\
\Q \Q $\tau$ \\
\Q \ALT \\
\Q \Q $\tau$; \\
\Q \Q self\_propagate$(n,X)$ \\
\Q \B{end alt} \\
\B{end process}
\end{tabular}
\end{small}
\end{center}
\caption{Equivalence checking for \I{TALLY} events (left) and \I{SELF\_PROPAGATE} events (right)}
\label{FIG-EC}
\end{figure}

        For instance, if one takes the (complex) LTS produced for a four-node
Algorand configuration, hides all events but \I{TALLY} (which is the event
used by each node to query its counter about vote results), renames in each
\I{TALLY (ID, $K_0$, $K_1$)} label the integer values $K_0$ and $K_1$
by some abstract constant $X$, and minimizes the result for branching
bisimulation, one obtains the (simple) LTS shown in Fig.~\ref{FIG-EC} (left).
This LTS is a bit too large to be checked visually, but one easily guesses it
is a cyclic four-dimensional hypercube that can be described by the LNT
code fragment shown on its right. The CADP tools are then used to formally
confirm this intuition.

        Similarly, if one takes the same (complex) LTS, hides all events
but \I{SELF\_PROPAGATE} (which is the event used by each node to communicate
its vote to its counter), renames in each \I{SELF\_PROPAGATE (ID, $B$)} label
the bit value $B$ by some abstract constant $X$, and minimizes the result
for branching bisimulation, one obtains the (not so simple) LTS shown in
Fig.~\ref{FIG-EC} (right). After many trials, it appears that this LTS
can be expressed by the LNT code fragment shown on its right (with the
definition of an auxiliary process $p$), which is confirmed by the CADP
tools.

\subsection{Model Checking}
\label{MODEL-CHECKING}

        Visual checking and equivalence checking may be intractable if the
model under verification is too large for bisimulation algorithms, or if the
property to be verified is complex and cannot be easily expressed as a simple
LTS. In both cases, model checking offers an alternative, in which the property
is specified as a temporal-logic formula given to an algorithm that computes
a Boolean truth value and a diagnostics (i.e., a counterexample) explaining
why the formula is true or false on the model.

        For instance, considering the \I{SET\_BIT} event (which is the event
introduced in the Algorand formal model to observe, at every step,  how each
node assigns its local bit $B$), event slicing for \I{SET\_BIT (ID, STEP, $B$)}
events does not give exploitable results. Even if one considers only one
chosen value of \I{ID}, hiding all other labels and renaming all bit values
$B$ by some abstract constant $X$, the resulting LTSs minimized for branching
bisimulation are small (25--50 states) but complex, with many (83\%) hidden
transitions (noted ``$\tau$'' as in CCS \cite{Milner-80}). It is thus easier
to express the properties that \I{SET\_BIT} events are expected to satisfy
using temporal-logic formulas rather than behavioural relations. An example
of such a formula, written in the MCL language~\cite{Mateescu-98-a}
\cite{Mateescu-Thivolle-08} and the SVL script language \cite{Garavel-Lang-01}
of CADP, is shown in Fig.~\ref{FIG-MC} --- notice the value-passing
capabilities of MCL, which are used to follow the successive values taken by
variables \I{ID} and \I{STEP}. Other MCL properties are given in
Annex~\ref{ANNEX-PROPERTIES}.

\begin{figure}
\begin{center}
\begin{small}
\begin{tabular}{l}
\B{property} P9b (MODEL) \\
\Q ``\I{for each \textcolor{blue}{ID} in {1...4}, after SET\_BIT (\textcolor{blue}{ID}, \textcolor{red}{STEP}, any), where \textcolor{red}{STEP} $<$ 2, it is inevitable to reach\/}'' \\
\Q ``\I{either RECEIVE\_BLOCK\_PROPOSAL (any) or SET\_BIT (\textcolor{blue}{ID}, \textcolor{red}{STEP} + 1, any) by following a\/}'' \\
\Q ``\I{path that contains neither RECEIVE\_BLOCK\_PROPOSAL (any) nor SET\_BIT (\textcolor{blue}{ID}, any, any)\/}'' \\
\B{is} \\
\Q ``\$\/MODEL.bcg'' $\mid$= \B{forall} \textcolor{blue}{ID}:NAT \B{among} \{1 ... 4\} . \\
\hspace*{30mm}\Q AFTER\_1\_WITHOUT\_2\_INEVITABLE\_3 ( \\
\hspace*{30mm}\Q\Q \{ SET\_BIT !\textcolor{blue}{ID} ?\textcolor{red}{STEP}:NAT ?\B{any} \B{where} \textcolor{red}{STEP} $<$ 2 \}, \\
\hspace*{30mm}\Q\Q \{ RECEIVE\_BLOCK\_PROPOSAL ?\B{any} \} \B{or} \{ SET\_BIT !\textcolor{blue}{ID} ?\B{any} ?\B{any} \}, \\
\hspace*{30mm}\Q\Q \{ RECEIVE\_BLOCK\_PROPOSAL ?\B{any} \} \B{or} \{ SET\_BIT !\textcolor{blue}{ID} !\textcolor{red}{STEP} + 1 ?\B{any} \})\:; \\
\Q -\,\!- \I{AFTER\_1\_WITHOUT\_2\_INEVITABLE\_3 (x,y,z) is a predefined macro of the MCL standard library} \\
\B{end property}
\end{tabular}
\end{small}
\end{center}
\caption{Model checking with a value-passing MCL formula embedded in an SVL script}
\label{FIG-MC}
\end{figure}


\section{Conclusion}
\label{CONCLUSION}

        Because they are designed to be minimal languages with tiny syntax
and concise semantics, old-school process calculi often lead to poorly
readable specifications --- very much like low-level code written in
assembly or bytecode languages.
        LNT supports the functional/recursive programming style of these
process calculi, but also brings useful features (imperative programming
style with direct assignments to variables, symmetric sequential composition,
and explicit loop operators) that, although present in the original version
of CSP \cite{Hoare-78}, are missing in CCS and most process calculi inspired
by CCS. Such a versatility in LNT allows models that are easier to read,
maintain, and evolve. Thanks to these features, LNT bridges the gap between
executable programs and formal models, by having a continuity in style
between these two classes of artefacts.

        To assess these claims, we experimented with a recent formal model
\cite{Esposito-Rossi-Bernardo-Fabris-Garavel-25} of the Algorand consensus
protocol \cite{Chen-Micali-19}. Starting from this initial version (967~lines
of LNT), we applied specific transformations to produce a simpler version
U0 (769 lines), which we further simplified, using four generic
transformations, to obtain four successive versions U1, U2, U3, and U4.
The latter version is concise (250 lines only), easily readable, and brings
new insight into the protocol.


        The four generic transformations presented in Sect.~\ref{MODULES}
to~\ref{LOOPS} are based on structural/algebraical properties of LNT and
are thus applicable to other case studies. They can be completed with other
generic transformations, such as those used in \cite{Garavel-Luttik-24}:
inline expansion of small auxiliary processes and flattening of nested
``\B{if-then-else}'' conditionals by adding ``\B{elsif}'' tests. All these
transformations can be done in a stepwise manner, checking that strong
or branching bisimulation is preserved at each step.

        We also verified various properties using visual checking,
equivalence checking, and model checking on the formal models of Algorand.
        The present work could be extended in at least four directions:

\begin{itemize}
        \item Keeping the model as it is, one could devise more correctness
properties to be verified. In particular, it would be challenging to specify
MCL formulas that globally observe the votes of all nodes and check whether
the decisions taken by consensus are really the expected ones.

        \item Because the model contains probabilistic choices, it would be
natural to perform probabilistic verification. This would require to complete
the model with the ``graded consensus'' phase (currently abstracted away) and
to better describe the notion of committees, either by making sure that the
size of committees is equal to $C$ or, at least, by considering only
committees whose size is larger than the threshold value $T$ and does not
have a too low occurrence probability.

        \item One could relax the synchrony assumption, materialized
in the model by the two synchronizations barriers \I{SYNC (BEGIN)} and
\I{SYNC (END)}, and adopt the ``substantially asynchronous'' setting of
\cite{Chen-Micali-19}, in which timeouts and overlapping rounds must be
taken into account.

        \item Finally, the model could shift from the BBA* protocol of
\cite{Chen-Micali-19} to its successor, the ABFT (Algorand Byzantine
Fault Tolerance) protocol, which is implemented and deployed, but
whose specifications~\cite{Algorand-specs-25} are dynamically evolving
over time, possibly causing moving-target issues for modellers.
\end{itemize}


\subsection*{Acknowledgements}

        We are grateful to Marco Bernardo, Andrea Esposito, and Francesco Pio
Rossi for undertaking a formal modelling of the Algorand consensus protocol
and for their patient explanations of its intricacies. We also thank
Pierre-Yves Piriou and the anonymous reviewers for their remarks
about the present article. The development of versions U0--U4 led to more
than~420 different models, which sometimes pushed the limits of LNT:
acknowledgements are due to Fr\'ed\'eric Lang and Wendelin Serwe for promptly
improving the LNT tools, and to Radu Mateescu for advising on the specification
 of MCL temporal-logic formulas.


\bibliographystyle{eptcs}
\bibliography{bibl_aux}


\appendix



\section{Preliminary Steps}

\subsection{Ambiguities in the Informal Description}
\label{ANNEX-AMBIGUITIES}

        We give two examples of ambiguities that we found in the original,
informal description \cite{Chen-Micali-19} and indicate which are the most
plausible interpretations:

\begin{itemize}
        \item The Algorand consensus protocol is temporally divided into
successive ``rounds'', and each round is divided into successive ``steps''.
One may wonder whether a committee is formed at each round, meaning that
the committee remains the same while a given block is examined
(interpretation~\#1) or at each step, meaning that several committees are
formed for the same block (interpretation~\#2).

        The difference between rounds and steps is clearly stated in
\cite{Chen-Micali-19}, e.g., page~160, where the variables $r$ and $s$ are
defined as follows: ``{\em $r \geq 0$ and $s \geq 1$ [denote] the current
round and the current step (in a given round)\/}''. In favor of
interpretation~\#1, page~156 defines: ``{\em a small set of selected verifiers,
noted $SV^r$, referred to as the committee\/}'', and this notation $SV^r$ is
used throughout pages 157--159, suggesting that the committee depends only on
round $r$. In favor of interpretation~\#2, page~159 formulates the concept of
``player replaceability'' and pages~160-161 introduce a new notation
$SV^{r,s}$ defined as ``{\em the set of verifiers of step $s$ of round $r$\/}.

        Following \cite{Esposito-Rossi-Bernardo-Fabris-Garavel-25}, all
formal models in LNT adopt interpretation \#2, meaning that a new committee
is formed at each step of each round.

        \item Another ambiguity concerns the amount of information passed
from a step to the next step. One may wonder whether this information is
empty (interpretation~\#1) or not (interpretation~\#2).

        In favor of interpretation~\#1, page~158 states, when defining
player replaceability: ``{\em no internal states need to be maintained by
a player from one step to another, and the protocol correctly reaches
consensus even if each step is executed by a totally new (independently and
randomly selected) set of players\/}''. In favor of interpretation~\#2,
pages 163--164 introduce, for each node $i$, a bit variable noted $b_i$,
the value of which is passed from the current step to the next step.

        Following \cite{Esposito-Rossi-Bernardo-Fabris-Garavel-25} again,
all formal models in LNT adopt interpretation \#2 and assume that each
node manages a state variable \I{B:BIT} that persists during an entire round.
\end{itemize}


\subsection{Simplifying Abstractions in the Formal Models}
\label{ANNEX-ABSTRACTIONS}

        The formal model presented in
\cite{Esposito-Rossi-Bernardo-Fabris-Garavel-25}, from which the LNT models
are derived, notably simplifies the Algorand
consensus protocol defined in \cite{Chen-Micali-19} by making various
abstractions intended to ease formal verification using finite-state
methods. We briefly summarize the main abstractions:

\begin{itemize}
        \item The number of nodes (i.e., users) in the network is fixed.
Nodes cannot join or leave the blockchain dynamically.

        \item The proportion of honest vs malicious nodes is fixed. A node is
either honest or malicious, and will remain so forever.

        \item Two particular configurations are studied: $A_{4,0}$ (a network
with four honest nodes) and $A_{2,2}$ (a network with two honest nodes and
two malicious nodes).

        \item Money and money transfers (e.g., rewards) are not modelled.
All nodes are assumed to have the same stake.

        \item The model focuses on the second phase ``BBA*'' (generalized
Boolean Byzantine Agreement) of the Algorand consensus protocol. The first
phase ``GC'' (Graded Consensus) is not described and its result is abstracted
away using a probabilistic choice.

        \item The role of leader nodes (elected during the GC phase) is not
modelled.

        \item Cryptographic aspects (credentials, public keys, secret keys,
signatures, etc.) are not described.

        \item Step numbers are abstracted away by using their residues
modulo~3 (noted $\equiv\!\!0$, $\equiv\!\!1$, and $\equiv\!\!2$).

        \item Probabilities are given constant values (e.g., 0.75 for the
probability to be in the committee and 0.7424 for the probability that an
elected leader is honest).

        \item The underlying network is assumed to be synchronous, whereas
Algorand was designed for a more permissive setting --- qualified as
``highly asynchronous'' or ``substantially asynchronous'' in
\cite{Chen-Micali-19}. The synchrony assumption greatly simplifies the formal
model, since each voting phase is delimited by two synchronization barriers
that concern all nodes in the network, so that delays and time bounds do not
have to be formalized.

        \item The Algorand notion of committee is loosely modelled.
Assuming that the network has $N$ nodes, votes take place within committees
of $C$ nodes, where $C$ is usually much smaller than $N$. But, since each
committee is formed probabilistically (using verifiable random functions),
its size may be different from $C$, as it determined by a binomial
distribution centered at $C$. Because of this, the formal model does not
only considers committees of size $C$, but all possible committees whose size
ranges between 0 and $N$, which somehow undermines the concept of committee.
\end{itemize}

\noindent
The formal model of \cite{Esposito-Rossi-Bernardo-Fabris-Garavel-25} also
introduces an attacker model (i.e., assumptions about malicious nodes) that
can be summarized as follows:

\begin{itemize}
        \item Malicious nodes cannot fork the blockchain (which is
probabilistically impossible by design in Algorand) but they seek to disrupt
it by preventing certain valid blocks from being committed.

        \item Malicious nodes influence the blockchain by setting their votes
to bit one (which means rejection).

        \item Malicious nodes share a private network. At the beginning of
each round, they may synchronize altogether on an event named \I{BOYCOTT}
and decide to attack the proposed block.
\end{itemize}


\subsection{Model-Specific Transformations}
\label{ANNEX-SPECIFIC}

The initial version U0 was produced by starting from the LNT model presented
in \cite[Sect. 4.2 and 4.3]{Esposito-Rossi-Bernardo-Fabris-Garavel-25},
to which the following (syntactic and semantic) transformations were applied:

\begin{enumerate}
        \item Modified spacing and indentation at various places.

        \item Renamed process $C$ to \I{COUNTER}.

        \item Added ``\B{ensure}'' post-condition in function $N$.

        \item Added constant function $C$ (committee size).

        \item Added constant function $H$ (number of honest nodes).

        \item Modified definition of function $T$ (threshold of votes).

        \item Added function \I{P\_V} (probability that a node is selected).

        \item Added function \I{P\_H} (probability that an elected leader is
honest).

        \item Replaced variable \I{P\_0} and its hard-coded value 0.7424 by
\I{P\_H}.

        \item Replaced variable \I{IN} and its hard-coded value 0.75 by
\I{P\_V}.

        \item Replaced ``action'' by ``event'' in comments. Shortened or
simplified certain comments.

        \item Added enumerated type \I{TAG} with two values (\I{BEGIN} and
\I{END}) to distinguish between the two \I{SYNC} events. Updated channel
\I{SYNCHRONIZE} accordingly.

        \item Tagged all local events \I{ADJUST\_BIT}, \I{ASK}, \I{COMPUTE\_BIT},
\I{P\_B}, \I{P\_IN}, \I{P\_OUT}, \I{REPLY}, \I{SELF\_PROPAGATE}, and
\I{SELF\_VERIFY} by giving them a first offer \I{ID}, which is the number of
the node that emits these events; these extra offers are needed to observe
the system and express properties to be verified. Updated the corresponing
channels \I{ASK}, \I{COMPUTE}, \I{PROBABILISTIC}, \I{REPLY},
\I{SELF\_PROPAGATE}, and \I{VERIFY} accordingly.

        \item Reduced the number of enumerated values in type \I{STEP} from
four to three by merging \I{S\_INIT} and \I{S\_TWO}. Turned \I{S\_INIT} into
a constant function equal to \I{S\_TWO}.

        \item Renamed the three constructors \I{S\_ZERO}, \I{S\_ONE}, and
\I{S\_TWO} of type \I{STEP} to 0, 1, and 2 for conciseness and for handling
them easily in MCL temporal-logic formulas.

        \item Introduced a new type \I{PID} that replaces type \I{NAT} for
node numbers; besides increased type safety, the definition of function
$N$ now derives from that of type \I{PID} and the ``\B{where}'' guard in
process \I{COUNTER} becomes simpler (no need to check min-max bounds).

        \item Replaced offers ``0 \B{of} \I{BIT}\/'' and ``1 \B{of} \I{BIT}\/''
by ``0'' and ``1'' in events \I{ASK} and \I{REPLY} because, as of CADP
version 2025-k, the LNT2LOTOS translator now exploits channel definitions to
resolve overloading.

        \item Moved comments related to \I{SELF\_PROPAGATE} and \I{PROPAGATE}
events from channel definitions to the \I{COUNTER} process.

        \item Added an ``\B{only if}'' guard in the \I{COUNTER} process
to allow compositional state-space generation.

        \item Moved \I{ADJUST\_BIT} events, which were placed before
bit assignments, after bit assignments.

        \item Moved \I{COMPUTE\_BIT} events, which were placed before
bit assignments, after bit assignments.

        \item Added a second offer \I{B:BIT} to both events \I{COMPUTE\_BIT}
and \I{ADJUST\_BIT}, so as to observe bit values after assignments.

        \item Introduced a new type \I{PROB}, distinct from reals, for
probability values. Updated functions \I{P\_H} and \I{P\_V} accordingly.

        \item Added a new function \I{1\_MINUS} to compute probabilities.
Simplified processes $N'$ and $N''$ by using this new function.

        \item Added a second parameter \I{P:PROB} to process $N'$. Modified
process $N''''$ to invoke $N'$ with probability value 0.5 (fair coin
tossing) rather than \I{P\_H}.

        \item Split \I{P\_B} events used to model probabilistic choices
into two distinct events \I{P\_ZERO} and \I{P\_ONE}, since the introduction
of probability 0.5 creates nondeterminism between both events \I{P\_B (ID, P)}
and \I{P\_B (ID, 1\_MINUS (P))}, making the model harder to observe and verify.
The names \I{P\_ZERO} and \I{P\_ONE} follow Algorand's conventions, where
consensus on bit zero (resp., one) means acceptance (resp., rejection) of the
proposed block.

        \item Shortened variable names \I{K\_0} and \I{K\_1} to \I{K0} and
\I{K1}, respectively.

        \item Merged both events \I{ASK} and \I{REPLY} into a single event
\I{TALLY}, which simplifies the internal protocol used by each node to query
its counter; from now on, each event \I{TALLY (ID, ?K0, ?K1)} replaces a former
sequence of four events \I{ASK (ID, 0)} $\rightarrow$ \I{REPLY (ID, ?K0)}
$\rightarrow$ \I{ASK (ID, 1)} $\rightarrow$ \I{REPLY (ID, ?K1)}, dividing by
four the number of reachable states, as observed in experiments.

        \item Removed \I{SELF\_VERIFY} events, which bring no useful
information, as every \I{SELF\_VERIFY (ID)} is immediately followed by
either a \I{P\_IN (ID, ...)} or \I{P\_OUT (ID, ...)} event, and reciprocally;
moreover, the two counter variables \I{K0} and \I{K1} can be reset by
\I{TALLY} events rather than \I{SELF\_VERIFY} events; such removal reduces
the number of reachable states by more than 25\%. Removed the \I{VERIFY}
channel too.

        \item Merged both events \I{ADJUST\_BIT (ID, B)} and
\I{COMPUTE\_BIT (ID, B)}, which now have roughly the same meaning, into
a single event \I{SET\_BIT (ID, B)}. Merged both channels \I{ADJUST} and
\I{COMPUTE} into a single channel \I{SET\_BIT}.

        \item Inserted a second offer \I{S:STEP} in \I{SET\_BIT (ID, S, B)}
events, so as to observe the current step of the protocol every time a bit
is assigned.

        \item Revised the attacker model: formerly, malicious nodes had
to synchronize altogether on the \I{BOYCOTT} event before attacking (i.e.,
tampering their votes to try rejecting the proposed block) and only
attacked when this synchronization succeeded; also, from a verification point
of view, if the \I{BOYCOTT} event was hidden, it was impossible to distinguish
between attacks and non-attacks; now, malicious nodes no longer synchronize
on \I{BOYCOTT} but attack if the contents of the proposed block matches a
given pattern (which we abstract away as a bit value). Created a new channel
\I{COMMIT} with an empty profile (no offers). Changed the channel of events
\I{COMMIT\_PROPOSED\_BLOCK} and \I{COMMIT\_EMPTY\_BLOCK} from \I{BLOCK}
to \I{COMMIT}. Added an offer \I{B:BIT} to event \I{RECEIVE\_BLOCK\_PROPOSAL}
and channel \I{BLOCK}. Modified process $N$ to trigger attacks when the
proposed block carries bit one. Removed event \I{BOYCOTT} and channel
\I{BOYCOTT}.
\end{enumerate}

\noindent
        Finally, to obtain fair statistics about the transformations proposed
in Sect.~\ref{TRADEOFFS}, we reverted, in the definition of process $N''$,
some of the code simplifications presented in Sect.~\ref{TRADEOFFS}, which
had been incorportated in~\cite{Esposito-Rossi-Bernardo-Fabris-Garavel-25}
by anticipation.



\section{Formal Model in LNT}
\label{ANNEX-MODEL}

This annex gives the LNT code of versions U2, U3, and U4. We do not
reproduce here versions U0 and U1, since their LNT code is nearly identical
(plus duplicated definitions and minus ``\I{M: MORALITY}'' parameters) to
that of version U2.


\subsection{Type Definitions}

This section reproduces the \T{TYPES.lnt} module of version U3, which defines
various types and their related functions. The \T{TYPES.lnt} module of version
U2 is identical to that of version U3, minus the ``\T{<>}'' function of type
\I{STEP}. The \T{TYPES.lnt} module of version U4 is identical to that of
version U3, minus the ``\T{=}'', ``\T{<>}'', \I{S\_INIT}, and \I{NEXT}
functions of type \I{STEP}.

\SPACING
\begin{lstlisting}[language=LNT]
module TYPES (BIT) is  -- the BIT type is imported from a predefined library

type PID is  -- node identifiers
   range 1...4 of NAT
   with <>, last
end type

type STEP is  -- steps of BBA* consensus protocol in [Chen-Micali-19]
   0,  -- congruent-to-0-modulo-3 step
   1,  -- congruent-to-1-modulo-3 step
   2   -- congruent-to-2-modulo-3 step
   with =, <>
end type

function S_INIT: STEP is   -- initialization step
   return 2
end function

function NEXT (S: STEP): STEP is  -- next step modulo 3
   case S in
      0 -> return 1
   |  1 -> return 2
   |  2 -> return 0
   end case
end function

type TAG is  -- synchronization tags
   BEGIN,
   END
end type

type MORALITY is  -- honesty values of nodes
   HONEST,
   MALICIOUS,
   DISGUISED  -- honest behavior of a malicious node
   with =
end type

type PROB is  -- probability
   P: REAL where 0.0 <= P and P <= 1.0
end type

function 1_MINUS (P: PROB): PROB is  -- complement of a probability
   return PROB (1.0 - REAL (P))
end function

end module
\end{lstlisting}
\SPACING


\subsection{Constant Definitions}

This subsection reproduces the \T{CONSTANT.lnt} module, which defines various
constants and is identical in versions U1, U2, U3, and U4.

\SPACING
\begin{lstlisting}[language=LNT]
module CONSTANTS (TYPES) is

function N: NAT is  -- number of nodes in the network (derived from type PID)
   ensure result > 0;
   return NAT (last of PID)
end function

function C: NAT is  -- committee size (parameter)
   ensure 0 < result and result <= N;
   return 3
end function

function H: NAT is  -- number of honest nodes (parameter)
   ensure 0 <= result and result <= N;
   return 2  -- 2 for configuration A_(2,2), 4 for configuration A_(4,0)
end function

function T: NAT is  -- threshold of votes needed to commit a block (parameter)
   ensure result >= (2 * (C + 1)) div 3;  -- i.e., (2 / 3) * C, rounded up
   return 2
end function

function P_V: PROB is  -- probability that a node is selected for the committee
   return PROB (REAL (C) / REAL (N))
end function

function P_H: PROB is  -- probability that an elected leader is honest
   var P: REAL in
      P := REAL (H) / REAL (N);  -- probability that a node is honest
      return PROB ((P ^ 2.0) * (1.0 + P - (P ^ 2.0)))  -- from Chen-Micali-19
   end var
end function

end module
\end{lstlisting}
\SPACING


\subsection{Channel Definitions}

This subsection reproduces the \T{CHANNELS.lnt} module, which defines various
channels (used to specify the types of LNT event parameters) and is identical
in versions U2, U3, and U4.

\SPACING
\begin{lstlisting}[language=LNT]
module CHANNELS (TYPES) is

channel BLOCK is  -- for block proposals
   (V: BIT)
end channel

channel COMMIT is  -- for commit events
   ()
end channel

channel SYNCHRONIZE is  -- for synchronization events
   (T: TAG)
end channel

channel SET_BIT is  -- for bit setting events
   (ID: PID, S: STEP, B: BIT)
end channel

channel PROPAGATE is  -- for bit propagation events
   (ID: PID, B: BIT)
end channel

channel SELF_PROPAGATE is  -- for bit self-propagation events
   (ID: PID, B: BIT)
end channel

channel TALLY is  -- for tally events
   (ID: PID, K0: NAT, K1: NAT)
end channel

channel PROBABILISTIC is  -- for probabilistic choices
   (ID: PID, P: PROB)
end channel

end module
\end{lstlisting}
\SPACING


\subsection{Main Process}

This subsection reproduces the \T{ALGORAND.lnt} module, which defines an
Algorand configuration with four nodes only and is identical in versions
U1, U2, U3, and U4.

\SPACING
\begin{lstlisting}[language=LNT]
module ALGORAND (NODE) is

process MAIN [RECEIVE_BLOCK_PROPOSAL: BLOCK, COMMIT_PROPOSED_BLOCK: COMMIT,
              COMMIT_EMPTY_BLOCK: COMMIT, SYNC: SYNCHRONIZE, SET_BIT: SET_BIT,
              PROPAGATE: PROPAGATE, SELF_PROPAGATE: SELF_PROPAGATE,
              TALLY: TALLY, P_IN, P_OUT, P_ZERO, P_ONE: PROBABILISTIC] is
   par RECEIVE_BLOCK_PROPOSAL, PROPAGATE, SYNC,
       COMMIT_PROPOSED_BLOCK, COMMIT_EMPTY_BLOCK in
      NODE [...] (1 of PID)
   ||
      NODE [...] (2 of PID)
   ||
      NODE [...] (3 of PID)
   ||
      NODE [...] (4 of PID)
   end par
end process

end module
\end{lstlisting}
\SPACING


\subsection{Counter Process (Version U3)}
\label{ANNEX-U3-COUNTER}

This subsection reproduces the \T{COUNTER.lnt} module that is specified in
the CCS-like style (using only action prefix and tail process recursion) and
is identical in versions U1, U2, and U3.

\SPACING
\begin{lstlisting}[language=LNT]
module COUNTER (TYPES, CONSTANTS, CHANNELS) is

process COUNTER [PROPAGATE: PROPAGATE, SELF_PROPAGATE: SELF_PROPAGATE,
                 TALLY: TALLY] (ID: PID, K0, K1: NAT) is
   var J: PID, B: BIT in
      alt
         only if K0 + K1 < N then
            alt
               -- propagation events go from other nodes to this counter
               PROPAGATE (?J, ?B) where J <> ID
            []
               -- self-propagation events go from each node to its counter
               SELF_PROPAGATE (ID, ?B)
            end alt;
            if B == 0 then
               COUNTER [...] (ID, K0 + 1, K1)
            else
               COUNTER [...] (ID, K0, K1 + 1)
            end if
         end if
      []
         TALLY (ID, K0, K1);
         COUNTER [...] (ID, 0, 0)
      end alt
   end var
end process

end module
\end{lstlisting}
\SPACING


\subsection{Counter Process (Version U4)}
\label{ANNEX-U4-COUNTER}

This subsection reproduces the \T{COUNTER.lnt} module that is specified in the
LNT imperative style (using assignments, symmetric sequential composition,
and loop operators).

\SPACING
\begin{lstlisting}[language=LNT]
module COUNTER (TYPES, CONSTANTS, CHANNELS) is

process COUNTER [PROPAGATE: PROPAGATE, SELF_PROPAGATE: SELF_PROPAGATE,
                 TALLY: TALLY] (ID: PID) is
   var B: BIT, J: PID, K0, K1: NAT in
      K0 := 0;
      K1 := 0;
      loop
         alt
            only if K0 + K1 < N then
               alt
                  -- propagation events go from other nodes to this counter
                  PROPAGATE (?J, ?B) where J <> ID
               []
                  -- self-propagation events go from each node to its counter
                  SELF_PROPAGATE (ID, ?B)
               end alt;
               if B == 0 then
                  K0 += 1
               else
                  K1 += 1
               end if
            end if
         []
            TALLY (ID, K0, K1);
            K0 := 0;
            K1 := 0
         end alt
      end loop
   end var
end process

end module
\end{lstlisting}
\SPACING


\subsection{Node Processes (Version U2)}
\label{ANNEX-U2-NODE}

This subsection reproduces the \T{NODE.lnt} module that is specified in the
CCS-like style (using only action prefix and tail process recursion).

\SPACING
\begin{lstlisting}[language=LNT]
module NODE (TYPES, CONSTANTS, CHANNELS, COUNTER) is

process NODE [RECEIVE_BLOCK_PROPOSAL: BLOCK, COMMIT_PROPOSED_BLOCK: COMMIT,
              COMMIT_EMPTY_BLOCK: COMMIT, SYNC: SYNCHRONIZE, SET_BIT: SET_BIT,
              PROPAGATE: PROPAGATE, SELF_PROPAGATE: SELF_PROPAGATE,
              TALLY: TALLY, P_IN, P_OUT, P_ZERO, P_ONE: PROBABILISTIC]
             (ID: PID) is
   par SELF_PROPAGATE, TALLY in
      N [...] (ID)
   ||
      COUNTER [...] (ID, 0, 0)
   end par
end process

process N [RECEIVE_BLOCK_PROPOSAL: BLOCK, COMMIT_PROPOSED_BLOCK: COMMIT,
           COMMIT_EMPTY_BLOCK: COMMIT, SYNC: SYNCHRONIZE, SET_BIT: SET_BIT,
           PROPAGATE: PROPAGATE, SELF_PROPAGATE: SELF_PROPAGATE,
           TALLY: TALLY, P_IN, P_OUT, P_ZERO, P_ONE: PROBABILISTIC]
          (ID: PID) is
   var V: BIT in
      RECEIVE_BLOCK_PROPOSAL (?V);
      if NAT (ID) <= H then
         N_PRIME [...] (ID, P_H, HONEST)
      elsif V = 1 then
         N_PRIME [...] (ID, P_H, MALICIOUS)
      else
         N_PRIME [...] (ID, P_H, DISGUISED)
      end if
   end var
end process

process N_PRIME [RECEIVE_BLOCK_PROPOSAL: BLOCK, COMMIT_PROPOSED_BLOCK: COMMIT,
                 COMMIT_EMPTY_BLOCK: COMMIT, SYNC: SYNCHRONIZE, SET_BIT: SET_BIT,
                 PROPAGATE: PROPAGATE, SELF_PROPAGATE: SELF_PROPAGATE,
                 TALLY: TALLY, P_IN, P_OUT, P_ZERO, P_ONE: PROBABILISTIC]
                (ID: PID, P: PROB, M: MORALITY) is
   alt
      P_ZERO (ID, P);
      N_SECOND [...] (ID, S_INIT, 0, M)
   []
      P_ONE (ID, 1_MINUS (P));
      N_SECOND [...] (ID, S_INIT, 1, M)
   end alt
end process

process N_SECOND [RECEIVE_BLOCK_PROPOSAL: BLOCK, COMMIT_PROPOSED_BLOCK: COMMIT,
                  COMMIT_EMPTY_BLOCK: COMMIT, SYNC: SYNCHRONIZE, SET_BIT: SET_BIT,
                  PROPAGATE: PROPAGATE, SELF_PROPAGATE: SELF_PROPAGATE,
                  TALLY: TALLY, P_IN, P_OUT, P_ZERO, P_ONE: PROBABILISTIC]
                 (ID: PID, S: STEP, in var B: BIT, M: MORALITY) is
   SET_BIT (ID, NEXT (S), B);
   B := B or BIT (M = MALICIOUS);
   SYNC (BEGIN);
   case S in
      (* S_INIT *) 2 ->  -- the LNT compiler wants a constructor here
         alt
            P_IN (ID, P_V);
            PROPAGATE (ID, B);
            SELF_PROPAGATE (ID, B);
            SYNC (END);
            N_THIRD [...] (ID, 0, M)
         []
            P_OUT (ID, 1_MINUS (P_V));
            SYNC (END);
            N_THIRD [...] (ID, 0, M)
         end alt
   |  0 ->
         alt
            P_IN (ID, P_V);
            PROPAGATE (ID, B);
            SELF_PROPAGATE (ID, B);
            SYNC (END);
            N_THIRD [...] (ID, 1, M)
         []
            P_OUT (ID, 1_MINUS (P_V));
            SYNC (END);
            N_THIRD [...] (ID, 1, M)
         end alt
   |  1 ->
         var K0, K1: NAT in
            alt
               P_IN (ID, P_V);
               PROPAGATE (ID, B);
               SELF_PROPAGATE (ID, B);
               SYNC (END);
               TALLY (ID, ?K0 where K0 <= N, ?K1 where K1 <= N);
               N_FOURTH [...] (ID, 2, K0, K1, M)
            []
               P_OUT (ID, 1_MINUS (P_V));
               SYNC (END);
               TALLY (ID, ?K0 where K0 <= N, ?K1 where K1 <= N);
               N_FOURTH [...] (ID, 2, K0, K1, M)
            end alt
         end var
   end case
end process

process N_THIRD [RECEIVE_BLOCK_PROPOSAL: BLOCK, COMMIT_PROPOSED_BLOCK: COMMIT,
                 COMMIT_EMPTY_BLOCK: COMMIT, SYNC: SYNCHRONIZE, SET_BIT: SET_BIT,
                 PROPAGATE: PROPAGATE, SELF_PROPAGATE: SELF_PROPAGATE,
                 TALLY: TALLY, P_IN, P_OUT, P_ZERO, P_ONE: PROBABILISTIC]
                (ID: PID, S: STEP, M: MORALITY) is
   require S = 0 or S = 1;
   var K0, K1: NAT in
      case S in
         0 ->
            TALLY (ID, ?K0 where K0 <= N, ?K1 where K1 <= N);
            if K0 >= T then
               COMMIT_PROPOSED_BLOCK;
               N [...] (ID)
            else
               N_FOURTH [...] (ID, 0, K0, K1, M)
            end if
      |  1 ->
            TALLY (ID, ?K0 where K0 <= N, ?K1 where K1 <= N);
            if K1 >= T then
               COMMIT_EMPTY_BLOCK;
               N [...] (ID)
            else
               N_FOURTH [...] (ID, 1, K0, K1, M)
            end if
      |  any ->
            raise UNEXPECTED
      end case
   end var
end process

process N_FOURTH [RECEIVE_BLOCK_PROPOSAL: BLOCK, COMMIT_PROPOSED_BLOCK: COMMIT,
                  COMMIT_EMPTY_BLOCK: COMMIT, SYNC: SYNCHRONIZE, SET_BIT: SET_BIT,
                  PROPAGATE: PROPAGATE, SELF_PROPAGATE: SELF_PROPAGATE,
                  TALLY: TALLY, P_IN, P_OUT, P_ZERO, P_ONE: PROBABILISTIC]
                 (ID: PID, S: STEP, K0, K1: NAT, M: MORALITY) is
   case S in
      0 ->
         if K1 >= T then
            N_SECOND [...] (ID, 0, 1, M)
         else
            N_SECOND [...] (ID, 0, 0, M)
         end if
   |  1 ->
         if K0 >= T then
            N_SECOND [...] (ID, 1, 0, M)
         else
            N_SECOND [...] (ID, 1, 1, M)
         end if
   |  2 ->
         if K0 >= T then
            N_SECOND [...] (ID, S_INIT, 0, M)
         elsif K1 >= T then
            N_SECOND [...] (ID, S_INIT, 1, M)
         else
            N_PRIME [...] (ID, 0.5 of PROB, M)
         end if
   end case
end process

end module
\end{lstlisting}
\SPACING


\subsection{Node Processes (Version U3)}
\label{ANNEX-U3-NODE}

This subsection reproduces the \T{NODE.lnt} module that is specified in an
halfway style combining elements from the CCS-like style (tail process
recursion) and the LNT imperative style (assignments and symmetric sequential
composition).

\SPACING
\begin{lstlisting}[language=LNT]
module NODE (TYPES, CONSTANTS, CHANNELS, COUNTER) is

process NODE [RECEIVE_BLOCK_PROPOSAL: BLOCK, COMMIT_PROPOSED_BLOCK: COMMIT,
              COMMIT_EMPTY_BLOCK: COMMIT, SYNC: SYNCHRONIZE, SET_BIT: SET_BIT,
              PROPAGATE: PROPAGATE, SELF_PROPAGATE: SELF_PROPAGATE,
              TALLY: TALLY, P_IN, P_OUT, P_ZERO, P_ONE: PROBABILISTIC]
             (ID: PID) is
   par SELF_PROPAGATE, TALLY in
      N [...] (ID)
   ||
      COUNTER [...] (ID, 0, 0)
   end par
end process

process N [RECEIVE_BLOCK_PROPOSAL: BLOCK, COMMIT_PROPOSED_BLOCK: COMMIT,
           COMMIT_EMPTY_BLOCK: COMMIT, SYNC: SYNCHRONIZE, SET_BIT: SET_BIT,
           PROPAGATE: PROPAGATE, SELF_PROPAGATE: SELF_PROPAGATE,
           TALLY: TALLY, P_IN, P_OUT, P_ZERO, P_ONE: PROBABILISTIC]
          (ID: PID) is
   var V: BIT, M: MORALITY in
      RECEIVE_BLOCK_PROPOSAL (?V);
      if NAT (ID) <= H then
         M := HONEST
      elsif V = 1 then
          M := MALICIOUS
      else
         M := DISGUISED
      end if;
      N_PRIME [...] (ID, P_H, M)
   end var
end process

process N_PRIME [RECEIVE_BLOCK_PROPOSAL: BLOCK, COMMIT_PROPOSED_BLOCK: COMMIT,
                 COMMIT_EMPTY_BLOCK: COMMIT, SYNC: SYNCHRONIZE, SET_BIT: SET_BIT,
                 PROPAGATE: PROPAGATE, SELF_PROPAGATE: SELF_PROPAGATE,
                 TALLY: TALLY, P_IN, P_OUT, P_ZERO, P_ONE: PROBABILISTIC]
                (ID: PID, P: PROB, M: MORALITY) is
   var B: BIT in
      alt
         P_ZERO (ID, P);
         B := 0
      []
         P_ONE (ID, 1_MINUS (P));
         B := 1
      end alt;
      N_SECOND [...] (ID, S_INIT, B, M)
   end var
end process

process N_SECOND [RECEIVE_BLOCK_PROPOSAL: BLOCK, COMMIT_PROPOSED_BLOCK: COMMIT,
                  COMMIT_EMPTY_BLOCK: COMMIT, SYNC: SYNCHRONIZE, SET_BIT: SET_BIT,
                  PROPAGATE: PROPAGATE, SELF_PROPAGATE: SELF_PROPAGATE,
                  TALLY: TALLY, P_IN, P_OUT, P_ZERO, P_ONE: PROBABILISTIC]
                 (ID: PID, S: STEP, in var B: BIT, M: MORALITY) is
   SET_BIT (ID, NEXT (S), B);
   B := B or BIT (M = MALICIOUS);
   SYNC (BEGIN);
   alt
      P_IN (ID, P_V);
      PROPAGATE (ID, B);
      SELF_PROPAGATE (ID, B)
   []
      P_OUT (ID, 1_MINUS (P_V))
   end alt;
   SYNC (END);
   case S in
      (* S_INIT *) 2 ->  -- the LNT compiler wants a constructor here
         N_THIRD [...] (ID, 0, M)
   |  0 ->
         N_THIRD [...] (ID, 1, M)
   |  1 ->
         var K0, K1: NAT in
            TALLY (ID, ?K0 where K0 <= N, ?K1 where K1 <= N);
            N_FOURTH [...] (ID, S_INIT, K0, K1, M)
         end var
   end case
end process

process N_THIRD [RECEIVE_BLOCK_PROPOSAL: BLOCK, COMMIT_PROPOSED_BLOCK: COMMIT,
                 COMMIT_EMPTY_BLOCK: COMMIT, SYNC: SYNCHRONIZE, SET_BIT: SET_BIT,
                 PROPAGATE: PROPAGATE, SELF_PROPAGATE: SELF_PROPAGATE,
                 TALLY: TALLY, P_IN, P_OUT, P_ZERO, P_ONE: PROBABILISTIC]
                (ID: PID, S: STEP, M: MORALITY) is
   require S = 0 or S = 1;
   var K0, K1: NAT in
      TALLY (ID, ?K0 where K0 <= N, ?K1 where K1 <= N);
      if (S = 0 and K0 >= T) or (S = 1 and K1 >= T) then
         if S = 0 then
            COMMIT_PROPOSED_BLOCK
         else
            COMMIT_EMPTY_BLOCK
         end if;
         N [...] (ID)
      else
         N_FOURTH [...] (ID, S, K0, K1, M)
      end if
   end var
end process

process N_FOURTH [RECEIVE_BLOCK_PROPOSAL: BLOCK, COMMIT_PROPOSED_BLOCK: COMMIT,
                  COMMIT_EMPTY_BLOCK: COMMIT, SYNC: SYNCHRONIZE, SET_BIT: SET_BIT,
                  PROPAGATE: PROPAGATE, SELF_PROPAGATE: SELF_PROPAGATE,
                  TALLY: TALLY, P_IN, P_OUT, P_ZERO, P_ONE: PROBABILISTIC]
                 (ID: PID, S: STEP, K0, K1: NAT, M: MORALITY) is
   var DONE: BOOL, B: BIT in
      DONE := false;
      B := 0;  -- dummy assignment to please the LNT compiler
      if S <> 0 then
         DONE := K0 >= T or S = 1;
         B := not (BIT (K0 >= T))
      end if;
      if not (DONE) and S <> 1 then
         DONE := K1 >= T or S = 0;
         B := BIT (K1 >= T)
      end if;
      if DONE then
         N_SECOND [...] (ID, S, B, M)
      else
         assert S = 2;
         N_PRIME [...] (ID, 0.5 of PROB, M)
      end if
   end var
end process

end module
\end{lstlisting}
\SPACING


\subsection{Node Processes (Version U4)}

This subsection reproduces the \T{NODE.lnt} module that is specified in the
LNT imperative style (using assignments, symmetric sequential composition,
and loop operators).

\SPACING
\begin{lstlisting}[language=LNT]
module NODE (TYPES, CONSTANTS, CHANNELS, COUNTER) is

process NODE [RECEIVE_BLOCK_PROPOSAL: BLOCK, COMMIT_PROPOSED_BLOCK: COMMIT,
              COMMIT_EMPTY_BLOCK: COMMIT, SYNC: SYNCHRONIZE, SET_BIT: SET_BIT,
              PROPAGATE: PROPAGATE, SELF_PROPAGATE: SELF_PROPAGATE,
              TALLY: TALLY, P_IN, P_OUT, P_ZERO, P_ONE: PROBABILISTIC]
             (ID: PID) is
   par SELF_PROPAGATE, TALLY in
      N [...] (ID)
   ||
      COUNTER [...] (ID)
   end par
end process

process N [RECEIVE_BLOCK_PROPOSAL: BLOCK, COMMIT_PROPOSED_BLOCK: COMMIT,
           COMMIT_EMPTY_BLOCK: COMMIT, SYNC: SYNCHRONIZE, SET_BIT: SET_BIT,
           PROPAGATE: PROPAGATE, SELF_PROPAGATE: SELF_PROPAGATE,
           TALLY: TALLY, P_IN, P_OUT, P_ZERO, P_ONE: PROBABILISTIC]
          (ID: PID) is
   var B, V: BIT, K0, K1: NAT, M: MORALITY in
      loop
         RECEIVE_BLOCK_PROPOSAL (?V);
         if NAT (ID) <= H then
            M := HONEST
         elsif V = 1 then
            M := MALICIOUS
         else
            M := DISGUISED
         end if;
         FLIP_COIN [...] (ID, 0, ?B, P_H);  -- result of graded consensus phase
         loop L in
            -- Coin-Fixed-To-0 step
            BROADCAST [...] (ID, B, M, ?K0, ?K1);
            if K0 >= T then
               COMMIT_PROPOSED_BLOCK;
               break L
            end if;
            FIX_COIN [...] (ID, 1, ?B, BIT (K1 >= T));
            -- Coin-Fixed-To-1 step
            BROADCAST [...] (ID, B, M, ?K0, ?K1);
            if K1 >= T then
               COMMIT_EMPTY_BLOCK;
               break L
            end if;
            FIX_COIN [...] (ID, 2, ?B, not (BIT (K0 >= T)));
            -- Coin-Genuinely-Flipped step
            BROADCAST [...] (ID, B, M, ?K0, ?K1);
            if K0 >= T then
               FIX_COIN [...] (ID, 0, ?B, 0)
            elsif K1 >= T then
               FIX_COIN [...] (ID, 0, ?B, 1)
            else
               FLIP_COIN [...] (ID, 0, ?B, 0.5 of PROB)  -- fair coin tossing
            end if
         end loop
      end loop
   end var
end process

process FIX_COIN [SET_BIT: SET_BIT]
                 (ID: PID, S: STEP, out var B: BIT, NEW_B: BIT) is
   B := NEW_B;
   SET_BIT (ID, S, B)
end process

process FLIP_COIN [SET_BIT: SET_BIT, P_ZERO, P_ONE: PROBABILISTIC]
                  (ID: PID, S: STEP, out var B: BIT, P: PROB) is
   alt
      P_ZERO (ID, P);
      B := 0
   []
      P_ONE (ID, 1_MINUS (P));
      B := 1
   end alt;
   SET_BIT (ID, S, B)
end process

process BROADCAST [SYNC: SYNCHRONIZE, PROPAGATE: PROPAGATE,
                   SELF_PROPAGATE: SELF_PROPAGATE, TALLY: TALLY,
                   P_IN, P_OUT: PROBABILISTIC]
                  (ID: PID, in var B: BIT, M: MORALITY, out K0, K1: NAT) is
   B := B or BIT (M = MALICIOUS);
   SYNC (BEGIN);
   alt
      P_IN (ID, P_V);
      PROPAGATE (ID, B);
      SELF_PROPAGATE (ID, B)
   []
      P_OUT (ID, 1_MINUS (P_V))
   end alt;
   SYNC (END);
   TALLY (ID, ?K0 where K0 <= N, ?K1 where K1 <= N)
end process

end module
\end{lstlisting}
\SPACING


\section{Verified Properties}
\label{ANNEX-PROPERTIES}

        The present annex lists a few properties, expressed in the SVL
language\footnote{\url{https://cadp.inria.fr/man/svl.html}}
\cite{Garavel-Lang-01}, which have been verified using CADP on both
configurations $A_{4,0}$ and $A_{2,2}$ of version U4.


\subsection{Absence of Deadlocks}

\begin{lstlisting}[language=SVL]
property P1 (MODEL)
   "there is no deadlock in $MODEL.bcg"
is
   deadlock of "$MODEL.bcg";
   expected FALSE
end property

\end{lstlisting}


\subsection{Properties of SYNC Events}

\begin{lstlisting}[language=SVL]
property P2 (MODEL)
   "the events SYNC (BEGIN) and SYNC (END) alternate forever in $MODEL.bcg"
is
   branching comparison
      hide all but SYNC in "$MODEL.bcg" == "SYNC.lnt";
   expected TRUE
end property
\end{lstlisting}

\noindent
where the auxiliary file \T{SYNC.lnt} contains the following definition:

\begin{lstlisting}[language=LNT]
module SYNC (TYPES, CHANNELS) is

process MAIN [SYNC: SYNCHRONIZE] is
   loop
      SYNC (BEGIN);
      SYNC (END)
   end loop
end process

end module
\end{lstlisting}


\subsection{Properties of TALLY Events}

\begin{lstlisting}[language=SVL,alsolanguage=MCL]
property P3a (MODEL)
   "the sum of both TALLY counters is always in the range 0...4"
is
   "$MODEL.bcg" |=
   not (POSSIBLE ({ TALLY ?any ?K0:nat ?K1:nat where K0 + K1 > 4 }));
   expected TRUE
end property
\end{lstlisting}

\begin{lstlisting}[language=SVL]
property P3b (MODEL)
   "the events TALLY (ID, X, X) form a cyclic 4-dimensional hypercube"
is
   branching comparison
    total rename "TALLY !\([0-9]\) !\([0-9]\) !\([0-9]\)" -> "TALLY !\1 !\"X\" !\"X\"" in
         hide all but TALLY in
            "$MODEL.bcg"
   ==
     "TALLY.lnt";
   expected TRUE
end property
\end{lstlisting}
\noindent
where the auxiliary file \T{TALLY.lnt} contains the following definition:

\begin{lstlisting}[language=LNT]
module TALLY (TYPES) is

process MAIN [TALLY: any] is
   loop
      par
         TALLY (1 of PID, "X", "X")
      ||
         TALLY (2 of PID, "X", "X")
      ||
         TALLY (3 of PID, "X", "X")
      ||
         TALLY (4 of PID, "X", "X")
      end par
   end loop
end process

end module

\end{lstlisting}


\subsection{Properties of PROPAGATE Events}

\begin{lstlisting}[language=SVL]
property P4a (MODEL)
   "the events PROPAGATE (1, any) form a simple looping automaton"
is
   branching comparison
      total hide all but "PROPAGATE !1 !.*" in "$MODEL.bcg"
   ==
      generation of "PROPAGATE_1.lnt";
   expected TRUE
end property
\end{lstlisting}
\noindent
where the auxiliary file \T{PROPAGATE\_1.lnt} contains the following definition:

\begin{lstlisting}[language=LNT]
module PROPAGATE_1 (TYPES, CHANNELS) is

process MAIN [PROPAGATE: PROPAGATE] is
   loop
      alt
         i;
         PROPAGATE (1 of PID, 0 of BIT)
      []
         i;
         PROPAGATE (1 of PID, 1 of BIT)
      end alt
   end loop
end process

end module

\end{lstlisting}

\begin{lstlisting}[language=SVL]
property P4b (MODEL)
   "the events PROPAGATE (any, X) form a complex looping automaton"
is
   branching comparison
      total rename "PROPAGATE !\([0-9]\) !.*" -> "PROPAGATE !\1 !\"X\"" in
         hide all but PROPAGATE in
            "$MODEL.bcg"
   ==
      generation of "PROPAGATE.lnt";
   expected TRUE
end property
\end{lstlisting}
\noindent
where the auxiliary file \T{PROPAGATE.lnt} contains the following definition:

\begin{lstlisting}[language=LNT]
module PROPAGATE (TYPES) is

process PART [PROPAGATE: any] (ID: PID) is
   alt
      i; PROPAGATE (ID, "X")
   []
      i
   end alt
end process

process MAIN [PROPAGATE: any] is
   loop
      par
         PART [...] (1 of PID)
      ||
         PART [...] (2 of PID)
      ||
         PART [...] (3 of PID)
      ||
         PART [...] (4 of PID)
      end par
   end loop
end process

end module

\end{lstlisting}


\subsection{Properties of SELF\_PROPAGATE Events}

\begin{lstlisting}[language=SVL]
property P5a (MODEL)
   "the events SELF_PROPAGATE (1, any) form a simple looping automaton"
is
   branching comparison
      total hide all but "SELF_PROPAGATE !1 !.*" in "$MODEL.bcg"
   ==
      generation of "SELF_PROPAGATE_1.lnt";
   expected TRUE
end property
\end{lstlisting}
\noindent
where the auxiliary file \T{SELF\_PROPAGATE\_1.lnt} contains the following
definition:

\begin{lstlisting}[language=LNT]
module SELF_PROPAGATE_1 (TYPES, CHANNELS) is

process MAIN [SELF_PROPAGATE: SELF_PROPAGATE] is
   loop
      var B: BIT in
         B := any BIT;
         i;
         SELF_PROPAGATE (1 of PID, B)
      end var
   end loop
end process

end module
\end{lstlisting}

\begin{lstlisting}[language=SVL]
property P5b (MODEL)
   "the events SELF_PROPAGATE (any, X) form a complex looping automaton"
is
   branching comparison
      total rename "SELF_PROPAGATE !\([0-9]\) !.*" -> "SELF_PROPAGATE !\1 !\"X\"" in
         hide all but SELF_PROPAGATE in
            "$MODEL.bcg"
   ==
      generation of "SELF_PROPAGATE.lnt";
   expected TRUE
end property
\end{lstlisting}
\noindent
where the auxiliary file \T{SELF\_PROPAGATE.lnt} contains the following
definition:

\begin{lstlisting}[language=LNT]
module SELF_PROPAGATE (TYPES) is

process PART [SELF_PROPAGATE: any] (ID: PID) is
   i;
   alt
      i; SELF_PROPAGATE (ID, "X")
   []
      i
   end alt
end process

process MAIN [SELF_PROPAGATE: any] is
   loop
      par
         PART [...] (1 of PID)
      ||
         PART [...] (2 of PID)
      ||
         PART [...] (3 of PID)
      ||
         PART [...] (4 of PID)
      end par
   end loop
end process

end module

\end{lstlisting}


\subsection{Properties of P\_IN and P\_OUT Events}

\begin{lstlisting}[language=SVL]
property P6a (MODEL)
   "the events P_IN (1, any) and P_OUT (1, any) form a simple looping automaton"
is
   branching comparison
      total hide all but "P_IN !1 !.*", "P_OUT !1 !.*" in "$MODEL.bcg"
   ==
      generation of "IN_OUT_1.lnt";
   expected TRUE
end property
\end{lstlisting}
\noindent
where the auxiliary file \T{IN\_OUT\_1.lnt} contains the following
definition:

\begin{lstlisting}[language=LNT]
module IN_OUT_1 (TYPES, CHANNELS, CONSTANTS) is

process MAIN [P_IN, P_OUT: PROBABILISTIC] is
   loop
      alt
         P_IN (1 of PID, P_V)
      []
         P_OUT (1 of PID, 1_MINUS (P_V))
      end alt
   end loop
end process

end module
\end{lstlisting}

\begin{lstlisting}[language=SVL]
property P6b (MODEL)
   "the events P_IN and P_OUT form a complex looping automaton"
is
   branching comparison
      hide all but P_IN, P_OUT in "$MODEL.bcg" == generation of "IN_OUT.lnt";
   expected TRUE
end property
\end{lstlisting}
\noindent
where the auxiliary file \T{IN\_OUT.lnt} contains the following
definition:

\begin{lstlisting}[language=LNT]
module IN_OUT (TYPES, CHANNELS, CONSTANTS) is

process ALT [P_IN, P_OUT: PROBABILISTIC] (ID: PID) is
   alt
      P_IN (ID, P_V)
   []
      P_OUT (ID, 1_MINUS (P_V))
   end alt
end process

process MAIN [P_IN, P_OUT: PROBABILISTIC] is
   loop
      par
         ALT [...] (1 of PID)
      ||
         ALT [...] (2 of PID)
      ||
         ALT [...] (3 of PID)
      ||
         ALT [...] (4 of PID)
      end par
   end loop
end process

end module
\end{lstlisting}


\subsection{Properties of P\_ZERO and P\_ONE Events}

\begin{lstlisting}[language=SVL]
property P7a (MODEL, H)  -- H: number of honest nodes in the configuration
   "the events P_ZERO (1, any) and P_ONE (1, any) form a simple looping automaton"
is
   branching comparison
      total hide all but "P_ZERO !1 !.*", "P_ONE !1 !.*" in "$MODEL.bcg"
   ==
      generation of "ZERO_ONE_1.lnt":MAIN [...] ($H);
   expected TRUE
end property
\end{lstlisting}
\noindent
where the auxiliary file \T{ZERO\_ONE\_1.lnt} contains the following
definition:

\begin{lstlisting}[language=LNT]
module ZERO_ONE_1 (TYPES, CHANNELS, CONSTANTS) is

process ALT [P_ZERO, P_ONE: PROBABILISTIC] (P: PROB) is
   alt
      P_ZERO (1 of PID, P)
   []
      P_ONE (1 of PID, 1_MINUS (P))
   end alt
end process

process MAIN [P_ZERO, P_ONE: PROBABILISTIC] (H: NAT) is
   var P: REAL, P_H: PROB in
      P := REAL (H) / REAL (N);
      P_H := PROB ((P ^ 2.0) * (1.0 + P - (P ^ 2.0)));
      ALT [...] (P_H);
      loop
         alt
            i;
            ALT [...] (0.5 of PROB)
         []
            i;
            ALT [...] (P_H)
         end alt
      end loop
   end var
end process

end module
\end{lstlisting}

\begin{lstlisting}[language=SVL]
property P7b (MODEL, H)  -- H: number of honest nodes in the configuration
   "the events P_ZERO and P_ONE form a complex looping automaton"
is
   branching comparison
      hide all but P_ZERO, P_ONE in "$MODEL.bcg"
   ==
      generation of "ZERO_ONE.lnt":MAIN [...] ($H);
   expected TRUE
end property
\end{lstlisting}
\noindent
where the auxiliary file \T{ZERO\_ONE.lnt} contains the following
definition:

\begin{lstlisting}[language=LNT]
module ZERO_ONE (TYPES, CHANNELS, CONSTANTS) is

process ALT [P_ZERO, P_ONE: PROBABILISTIC] (ID: PID, P: PROB) is
   alt
      P_ZERO (ID, P)
   []
      P_ONE (ID, 1_MINUS (P))
   end alt
end process

process PART [P_ZERO, P_ONE: PROBABILISTIC] (P: PROB) is
   par
      ALT [...] (1 of PID, P)
   ||
      ALT [...] (2 of PID, P)
   ||
      ALT [...] (3 of PID, P)
   ||
      ALT [...] (4 of PID, P)
   end par
end process

process MAIN [P_ZERO, P_ONE: PROBABILISTIC] (H: NAT) is
   var P: REAL, P_H: PROB in
      P := REAL (H) / REAL (N);
      P_H := PROB ((P ^ 2.0) * (1.0 + P - (P ^ 2.0)));
      PART [...] (P_H);
      loop
         alt
            i;
            PART [...] (0.5 of PROB)
         []
            i;
            PART [...] (P_H)
         end alt
      end loop
   end var
end process

end module
\end{lstlisting}


\subsection{Properties of BLOCK-related Events}

\begin{lstlisting}[language=SVL]
property P8a (MODEL)
   "observed alone, the events RECEIVE_BLOCK_PROPOSAL, COMMIT_PROPOSED_BLOCK,"
   "and COMMIT_EMPTY_BLOCK show the absence or presence of successful attacks"
is
   "${MODEL}_min.bcg" = hide all but RECEIVE_BLOCK_PROPOSAL,
                        COMMIT_PROPOSED_BLOCK, COMMIT_EMPTY_BLOCK in
                          "$MODEL.bcg";
   branching comparison "${MODEL}_min.bcg" == generation of "OK.lnt";
   result R1;
%  if [ $R1 = "TRUE" ]
%  then
%     echo "OK"
%     return
%  fi
   branching comparison "${MODEL}_min.bcg" == generation of "KO.lnt";
   result R2;
%  if [ $R2 = "TRUE" ]
%  then
%     echo "KO"
%     return
%  fi
end property
\end{lstlisting}
\noindent
where the auxiliary file \T{OK.lnt} contains the following definition:

\begin{lstlisting}[language=LNT]
module OK (TYPES, CHANNELS) is

process MAIN [RECEIVE_BLOCK_PROPOSAL: BLOCK, COMMIT_PROPOSED_BLOCK: COMMIT,
              COMMIT_EMPTY_BLOCK: COMMIT] is
   loop
      RECEIVE_BLOCK_PROPOSAL (?any BIT);
      alt
         i;
         COMMIT_PROPOSED_BLOCK
      []
         i;
         COMMIT_EMPTY_BLOCK
      end alt
   end loop
end process

end module
\end{lstlisting}
\noindent
and the auxiliary file \T{KO.lnt} contains the following definition:

\begin{lstlisting}[language=LNT]
module KO (TYPES, CHANNELS) is

process MAIN [RECEIVE_BLOCK_PROPOSAL: BLOCK, COMMIT_PROPOSED_BLOCK: COMMIT,
              COMMIT_EMPTY_BLOCK: COMMIT] is
   loop
      alt
         RECEIVE_BLOCK_PROPOSAL (0 of BIT);
         alt
            i;
            COMMIT_PROPOSED_BLOCK
         []
            i;
            COMMIT_EMPTY_BLOCK
         end alt
      []
         RECEIVE_BLOCK_PROPOSAL (1 of BIT);
         COMMIT_EMPTY_BLOCK
      end alt
   end loop
end process

end module
\end{lstlisting}


\subsection{Properties of SET\_BIT Events}

\begin{lstlisting}[language=SVL,alsolanguage=MCL]
property P9a (MODEL)
   "after RECEIVE_BLOCK_PROPOSAL (any), for each ID in {1...4}, it is inevitable"
   "to reach a SET_BIT (ID, 0, any) event by following a path that contains"
   "neither RECEIVE_BLOCK_PROPOSAL (any) nor SET_BIT (ID, 1 or 2, any)"
is
   "$MODEL.bcg" |=
   forall ID:NAT among {1 ... 4} .
      AFTER_1_WITHOUT_2_INEVITABLE_3 (
         { RECEIVE_BLOCK_PROPOSAL ?any },
         { RECEIVE_BLOCK_PROPOSAL ?any } or { SET_BIT !ID ?any ?any },
         { SET_BIT !ID !0 ?any });
end property
\end{lstlisting}

\begin{lstlisting}[language=SVL,alsolanguage=MCL]
property P9b (MODEL)
   "for each ID in {1...4}, after SET_BIT (ID, STEP, any), where STEP < 2,"
   "it is inevitable to reach either SET_BIT (ID, STEP + 1, any) or"
   "RECEIVE_BLOCK_PROPOSAL (any) by following a path that contains neither"
   "RECEIVE_BLOCK_PROPOSAL (any) nor SET_BIT (ID, any, any)"
is
   "$MODEL.bcg" |=
   forall ID:NAT among {1 ... 4} .
      AFTER_1_WITHOUT_2_INEVITABLE_3 (
         { SET_BIT !ID ?STEP:NAT ?any where STEP < 2 },
         { RECEIVE_BLOCK_PROPOSAL ?any } or { SET_BIT !ID ?any ?any },
         { RECEIVE_BLOCK_PROPOSAL ?any } or { SET_BIT !ID !STEP + 1 ?any });
end property
\end{lstlisting}

\begin{lstlisting}[language=SVL,alsolanguage=MCL]
property P9c (MODEL)
   "for each ID in {1...4}, after SET_BIT (ID, 2, any), it is inevitable"
   "to reach either SET_BIT (ID, 0, any) by following a path that contains"
   "neither RECEIVE_BLOCK_PROPOSAL (any) nor SET_BIT (ID, any, any)"
is
   "$MODEL.bcg" |=
   forall ID:NAT among {1 ... 4} .
      AFTER_1_WITHOUT_2_INEVITABLE_3 (
         { SET_BIT !ID !2 ?any },
         { RECEIVE_BLOCK_PROPOSAL ?any } or { SET_BIT !ID ?any ?any },
         { SET_BIT !ID !0 ?any });
end property
\end{lstlisting}

\end{document}

%% file: fig_u0.tex
\begin{picture}(0,0)%
\includegraphics{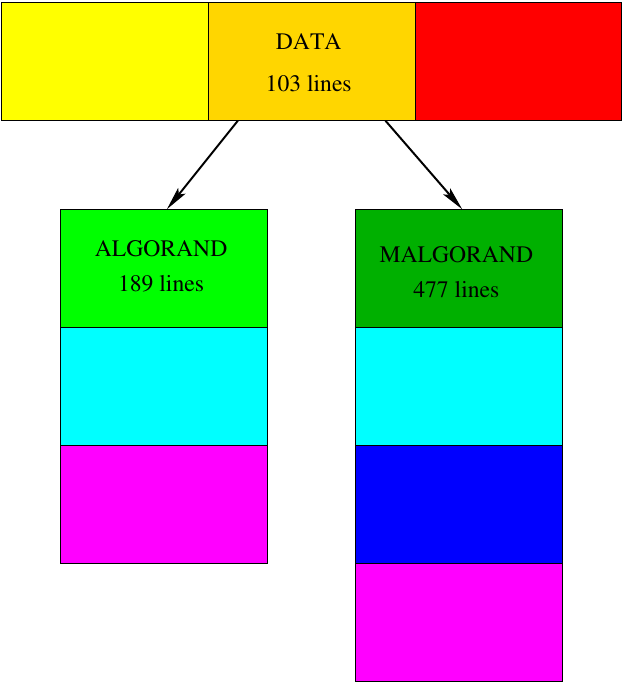}%
\end{picture}%
\setlength{\unitlength}{4144sp}%
\begingroup\makeatletter\ifx\SetFigFont\undefined%
\gdef\SetFigFont#1#2#3#4#5{%
  \reset@font\fontsize{#1}{#2pt}%
  \fontfamily{#3}\fontseries{#4}\fontshape{#5}%
  \selectfont}%
\fi\endgroup%
\begin{picture}(4749,5199)(1789,-6373)
\end{picture}%

%% file: fig_u1.tex
\begin{picture}(0,0)%
\includegraphics{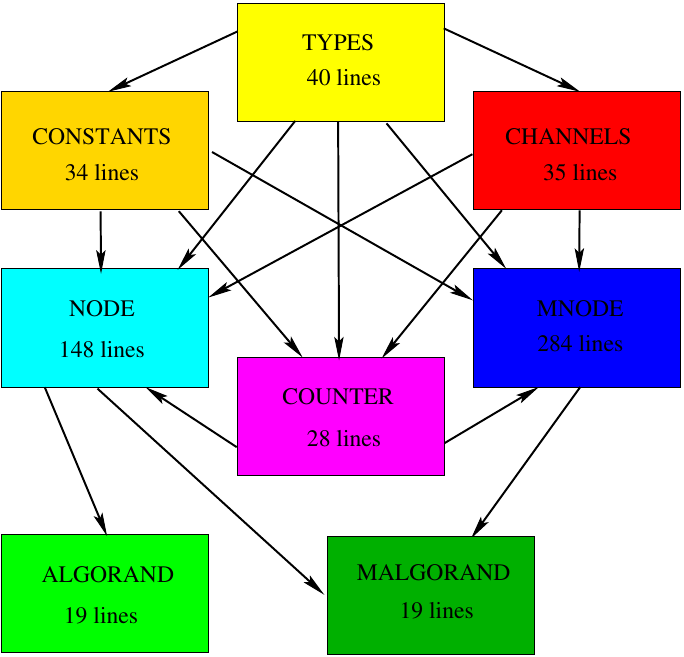}%
\end{picture}%
\setlength{\unitlength}{4144sp}%
\begingroup\makeatletter\ifx\SetFigFont\undefined%
\gdef\SetFigFont#1#2#3#4#5{%
  \reset@font\fontsize{#1}{#2pt}%
  \fontfamily{#3}\fontseries{#4}\fontshape{#5}%
  \selectfont}%
\fi\endgroup%
\begin{picture}(5199,4989)(1339,-6388)
\end{picture}%

%% file: fig_u2.tex
\begin{picture}(0,0)%
\includegraphics{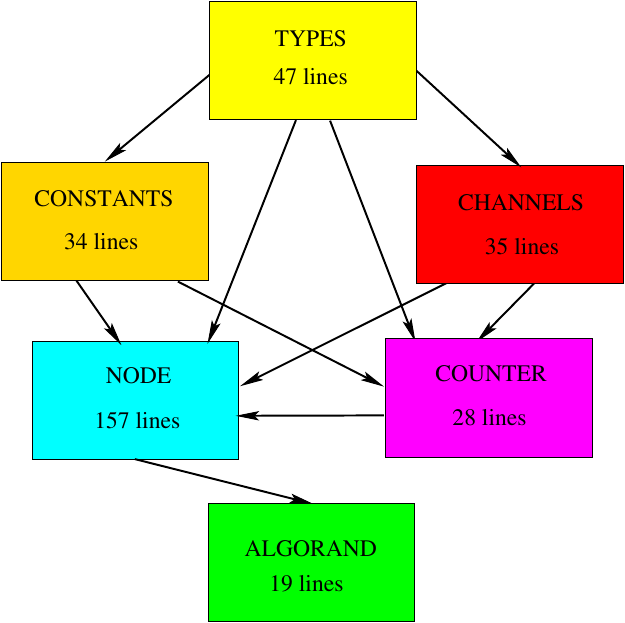}%
\end{picture}%
\setlength{\unitlength}{4144sp}%
\begingroup\makeatletter\ifx\SetFigFont\undefined%
\gdef\SetFigFont#1#2#3#4#5{%
  \reset@font\fontsize{#1}{#2pt}%
  \fontfamily{#3}\fontseries{#4}\fontshape{#5}%
  \selectfont}%
\fi\endgroup%
\begin{picture}(4762,4745)(1565,-6373)
\end{picture}%

%% file: fig_u4a.tex
\begin{picture}(0,0)%
\includegraphics{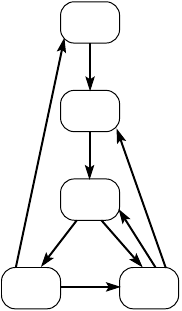}%
\end{picture}%
\setlength{\unitlength}{4144sp}%
\begingroup\makeatletter\ifx\SetFigFont\undefined%
\gdef\SetFigFont#1#2#3#4#5{%
  \reset@font\fontsize{#1}{#2pt}%
  \fontfamily{#3}\fontseries{#4}\fontshape{#5}%
  \selectfont}%
\fi\endgroup%
\begin{picture}(1374,2364)(664,-1963)
\put(1801,-1861){\makebox(0,0)[b]{\smash{{\SetFigFont{12}{14.4}{\familydefault}{\mddefault}{\updefault}{\color[rgb]{0,0,0}$N''''$}%
}}}}
\put(901,-1861){\makebox(0,0)[b]{\smash{{\SetFigFont{12}{14.4}{\familydefault}{\mddefault}{\updefault}{\color[rgb]{0,0,0}$N'''$}%
}}}}
\put(1351,-511){\makebox(0,0)[b]{\smash{{\SetFigFont{12}{14.4}{\familydefault}{\mddefault}{\updefault}{\color[rgb]{0,0,0}$N'$}%
}}}}
\put(1351,-1186){\makebox(0,0)[b]{\smash{{\SetFigFont{12}{14.4}{\familydefault}{\mddefault}{\updefault}{\color[rgb]{0,0,0}$N''$}%
}}}}
\put(1351,164){\makebox(0,0)[b]{\smash{{\SetFigFont{12}{14.4}{\familydefault}{\mddefault}{\updefault}{\color[rgb]{0,0,0}$N$}%
}}}}
\end{picture}%

%% file: fig_u4b.tex
\begin{picture}(0,0)%
\includegraphics{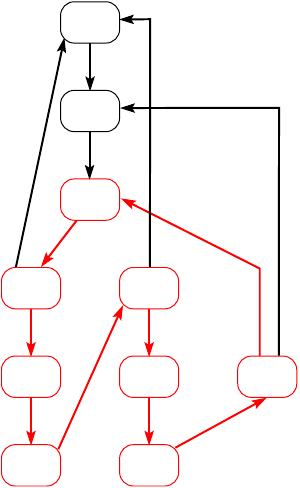}%
\end{picture}%
\setlength{\unitlength}{4144sp}%
\begingroup\makeatletter\ifx\SetFigFont\undefined%
\gdef\SetFigFont#1#2#3#4#5{%
  \reset@font\fontsize{#1}{#2pt}%
  \fontfamily{#3}\fontseries{#4}\fontshape{#5}%
  \selectfont}%
\fi\endgroup%
\begin{picture}(2274,3714)(664,-3313)
\put(1801,-1861){\makebox(0,0)[b]{\smash{{\SetFigFont{12}{14.4}{\familydefault}{\mddefault}{\updefault}{\color[rgb]{0,0,0}$N'''_1$}%
}}}}
\put(901,-1861){\makebox(0,0)[b]{\smash{{\SetFigFont{12}{14.4}{\familydefault}{\mddefault}{\updefault}{\color[rgb]{0,0,0}$N'''_0$}%
}}}}
\put(1351,-511){\makebox(0,0)[b]{\smash{{\SetFigFont{12}{14.4}{\familydefault}{\mddefault}{\updefault}{\color[rgb]{0,0,0}$N'$}%
}}}}
\put(1351,-1186){\makebox(0,0)[b]{\smash{{\SetFigFont{12}{14.4}{\familydefault}{\mddefault}{\updefault}{\color[rgb]{0,0,0}$N''_2$}%
}}}}
\put(1351,164){\makebox(0,0)[b]{\smash{{\SetFigFont{12}{14.4}{\familydefault}{\mddefault}{\updefault}{\color[rgb]{0,0,0}$N$}%
}}}}
\put(901,-2536){\makebox(0,0)[b]{\smash{{\SetFigFont{12}{14.4}{\familydefault}{\mddefault}{\updefault}{\color[rgb]{0,0,0}$N''''_0$}%
}}}}
\put(1846,-2536){\makebox(0,0)[b]{\smash{{\SetFigFont{12}{14.4}{\familydefault}{\mddefault}{\updefault}{\color[rgb]{0,0,0}$N''''_1$}%
}}}}
\put(901,-3211){\makebox(0,0)[b]{\smash{{\SetFigFont{12}{14.4}{\familydefault}{\mddefault}{\updefault}{\color[rgb]{0,0,0}$N''_0$}%
}}}}
\put(1801,-3211){\makebox(0,0)[b]{\smash{{\SetFigFont{12}{14.4}{\familydefault}{\mddefault}{\updefault}{\color[rgb]{0,0,0}$N''_1$}%
}}}}
\put(2701,-2536){\makebox(0,0)[b]{\smash{{\SetFigFont{12}{14.4}{\familydefault}{\mddefault}{\updefault}{\color[rgb]{0,0,0}$N''''_2$}%
}}}}
\end{picture}%

%% file: fig_u4c.tex
\begin{picture}(0,0)%
\includegraphics{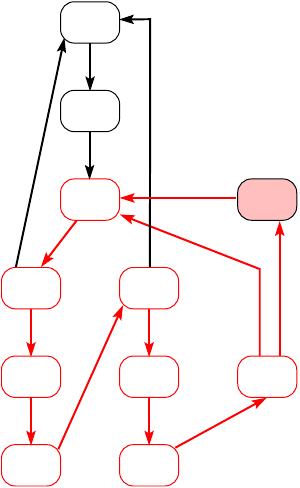}%
\end{picture}%
\setlength{\unitlength}{4144sp}%
\begingroup\makeatletter\ifx\SetFigFont\undefined%
\gdef\SetFigFont#1#2#3#4#5{%
  \reset@font\fontsize{#1}{#2pt}%
  \fontfamily{#3}\fontseries{#4}\fontshape{#5}%
  \selectfont}%
\fi\endgroup%
\begin{picture}(2274,3714)(664,-3313)
\put(1801,-1861){\makebox(0,0)[b]{\smash{{\SetFigFont{12}{14.4}{\familydefault}{\mddefault}{\updefault}{\color[rgb]{0,0,0}$N'''_1$}%
}}}}
\put(901,-1861){\makebox(0,0)[b]{\smash{{\SetFigFont{12}{14.4}{\familydefault}{\mddefault}{\updefault}{\color[rgb]{0,0,0}$N'''_0$}%
}}}}
\put(1351,-511){\makebox(0,0)[b]{\smash{{\SetFigFont{12}{14.4}{\familydefault}{\mddefault}{\updefault}{\color[rgb]{0,0,0}$N'$}%
}}}}
\put(1351,-1186){\makebox(0,0)[b]{\smash{{\SetFigFont{12}{14.4}{\familydefault}{\mddefault}{\updefault}{\color[rgb]{0,0,0}$N''_2$}%
}}}}
\put(1351,164){\makebox(0,0)[b]{\smash{{\SetFigFont{12}{14.4}{\familydefault}{\mddefault}{\updefault}{\color[rgb]{0,0,0}$N$}%
}}}}
\put(901,-2536){\makebox(0,0)[b]{\smash{{\SetFigFont{12}{14.4}{\familydefault}{\mddefault}{\updefault}{\color[rgb]{0,0,0}$N''''_0$}%
}}}}
\put(1846,-2536){\makebox(0,0)[b]{\smash{{\SetFigFont{12}{14.4}{\familydefault}{\mddefault}{\updefault}{\color[rgb]{0,0,0}$N''''_1$}%
}}}}
\put(901,-3211){\makebox(0,0)[b]{\smash{{\SetFigFont{12}{14.4}{\familydefault}{\mddefault}{\updefault}{\color[rgb]{0,0,0}$N''_0$}%
}}}}
\put(1801,-3211){\makebox(0,0)[b]{\smash{{\SetFigFont{12}{14.4}{\familydefault}{\mddefault}{\updefault}{\color[rgb]{0,0,0}$N''_1$}%
}}}}
\put(2701,-2536){\makebox(0,0)[b]{\smash{{\SetFigFont{12}{14.4}{\familydefault}{\mddefault}{\updefault}{\color[rgb]{0,0,0}$N''''_2$}%
}}}}
\put(2701,-1186){\makebox(0,0)[b]{\smash{{\SetFigFont{12}{14.4}{\familydefault}{\mddefault}{\updefault}{\color[rgb]{0,0,0}$N'$}%
}}}}
\end{picture}%

%% file: fig_vc_d1.tex
\begin{picture}(0,0)%
\includegraphics{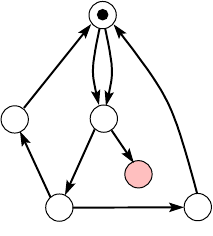}%
\end{picture}%
\setlength{\unitlength}{4144sp}%
\begingroup\makeatletter\ifx\SetFigFont\undefined%
\gdef\SetFigFont#1#2#3#4#5{%
  \reset@font\fontsize{#1}{#2pt}%
  \fontfamily{#3}\fontseries{#4}\fontshape{#5}%
  \selectfont}%
\fi\endgroup%
\begin{picture}(1619,1789)(2481,-1397)
\put(3468,-1337){\makebox(0,0)[b]{\smash{{\SetFigFont{12}{14.4}{\familydefault}{\mddefault}{\updefault}{\color[rgb]{0,0,0}$\tau$}%
}}}}
\put(2564,-947){\makebox(0,0)[b]{\smash{{\SetFigFont{12}{14.4}{\familydefault}{\mddefault}{\updefault}{\color[rgb]{0,0,0}$\tau$}%
}}}}
\put(3565,-745){\makebox(0,0)[b]{\smash{{\SetFigFont{12}{14.4}{\familydefault}{\mddefault}{\updefault}{\color[rgb]{0,0,0}$\tau$}%
}}}}
\put(2985,-745){\makebox(0,0)[b]{\smash{{\SetFigFont{12}{14.4}{\familydefault}{\mddefault}{\updefault}{\color[rgb]{0,0,0}$\tau$}%
}}}}
\put(2777,-136){\makebox(0,0)[b]{\smash{{\SetFigFont{12}{14.4}{\familydefault}{\mddefault}{\updefault}{\color[rgb]{0,0,0}$c$}%
}}}}
\put(3057,-247){\makebox(0,0)[b]{\smash{{\SetFigFont{12}{14.4}{\familydefault}{\mddefault}{\updefault}{\color[rgb]{0,0,0}$r_0$}%
}}}}
\put(3934,-462){\makebox(0,0)[b]{\smash{{\SetFigFont{12}{14.4}{\familydefault}{\mddefault}{\updefault}{\color[rgb]{0,0,0}$e$}%
}}}}
\put(3445,-259){\makebox(0,0)[b]{\smash{{\SetFigFont{12}{14.4}{\familydefault}{\mddefault}{\updefault}{\color[rgb]{0,0,0}$r_1$}%
}}}}
\end{picture}%

%% file: fig_vc_d2.tex
\begin{picture}(0,0)%
\includegraphics{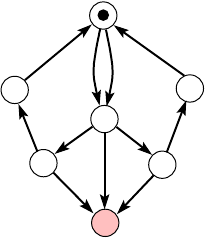}%
\end{picture}%
\setlength{\unitlength}{4144sp}%
\begingroup\makeatletter\ifx\SetFigFont\undefined%
\gdef\SetFigFont#1#2#3#4#5{%
  \reset@font\fontsize{#1}{#2pt}%
  \fontfamily{#3}\fontseries{#4}\fontshape{#5}%
  \selectfont}%
\fi\endgroup%
\begin{picture}(1559,1802)(2476,-1410)
\put(3769,-17){\makebox(0,0)[b]{\smash{{\SetFigFont{12}{14.4}{\familydefault}{\mddefault}{\updefault}{\color[rgb]{0,0,0}$e$}%
}}}}
\put(2992,-619){\makebox(0,0)[b]{\smash{{\SetFigFont{12}{14.4}{\familydefault}{\mddefault}{\updefault}{\color[rgb]{0,0,0}$\tau$}%
}}}}
\put(3541,-638){\makebox(0,0)[b]{\smash{{\SetFigFont{12}{14.4}{\familydefault}{\mddefault}{\updefault}{\color[rgb]{0,0,0}$\tau$}%
}}}}
\put(3972,-612){\makebox(0,0)[b]{\smash{{\SetFigFont{12}{14.4}{\familydefault}{\mddefault}{\updefault}{\color[rgb]{0,0,0}$\tau$}%
}}}}
\put(3674,-1137){\makebox(0,0)[b]{\smash{{\SetFigFont{12}{14.4}{\familydefault}{\mddefault}{\updefault}{\color[rgb]{0,0,0}$\tau$}%
}}}}
\put(2878,-1120){\makebox(0,0)[b]{\smash{{\SetFigFont{12}{14.4}{\familydefault}{\mddefault}{\updefault}{\color[rgb]{0,0,0}$\tau$}%
}}}}
\put(3182,-906){\makebox(0,0)[b]{\smash{{\SetFigFont{12}{14.4}{\familydefault}{\mddefault}{\updefault}{\color[rgb]{0,0,0}$\tau$}%
}}}}
\put(2812, -7){\makebox(0,0)[b]{\smash{{\SetFigFont{12}{14.4}{\familydefault}{\mddefault}{\updefault}{\color[rgb]{0,0,0}$c$}%
}}}}
\put(3066,-221){\makebox(0,0)[b]{\smash{{\SetFigFont{12}{14.4}{\familydefault}{\mddefault}{\updefault}{\color[rgb]{0,0,0}$r_0$}%
}}}}
\put(3460,-230){\makebox(0,0)[b]{\smash{{\SetFigFont{12}{14.4}{\familydefault}{\mddefault}{\updefault}{\color[rgb]{0,0,0}$r_1$}%
}}}}
\put(2550,-644){\makebox(0,0)[b]{\smash{{\SetFigFont{12}{14.4}{\familydefault}{\mddefault}{\updefault}{\color[rgb]{0,0,0}$\tau$}%
}}}}
\end{picture}%

%% file: fig_vc_ok.tex
\begin{picture}(0,0)%
\includegraphics{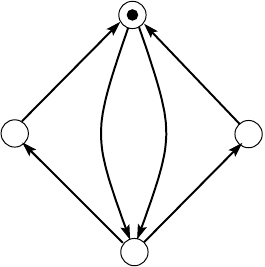}%
\end{picture}%
\setlength{\unitlength}{4144sp}%
\begingroup\makeatletter\ifx\SetFigFont\undefined%
\gdef\SetFigFont#1#2#3#4#5{%
  \reset@font\fontsize{#1}{#2pt}%
  \fontfamily{#3}\fontseries{#4}\fontshape{#5}%
  \selectfont}%
\fi\endgroup%
\begin{picture}(2006,2029)(2262,-2309)
\put(3698,-1343){\makebox(0,0)[b]{\smash{{\SetFigFont{12}{14.4}{\familydefault}{\mddefault}{\updefault}{\color[rgb]{0,0,0}$r_1$}%
}}}}
\put(2881,-1343){\makebox(0,0)[b]{\smash{{\SetFigFont{12}{14.4}{\familydefault}{\mddefault}{\updefault}{\color[rgb]{0,0,0}$r_0$}%
}}}}
\put(2552,-1797){\makebox(0,0)[b]{\smash{{\SetFigFont{12}{14.4}{\familydefault}{\mddefault}{\updefault}{\color[rgb]{0,0,0}$\tau$}%
}}}}
\put(3969,-1779){\makebox(0,0)[b]{\smash{{\SetFigFont{12}{14.4}{\familydefault}{\mddefault}{\updefault}{\color[rgb]{0,0,0}$\tau$}%
}}}}
\put(3826,-781){\makebox(0,0)[b]{\smash{{\SetFigFont{12}{14.4}{\familydefault}{\mddefault}{\updefault}{\color[rgb]{0,0,0}$e$}%
}}}}
\put(2701,-781){\makebox(0,0)[b]{\smash{{\SetFigFont{12}{14.4}{\familydefault}{\mddefault}{\updefault}{\color[rgb]{0,0,0}$c$}%
}}}}
\end{picture}%

%% file: fig_vc_ko.tex
\begin{picture}(0,0)%
\includegraphics{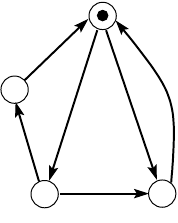}%
\end{picture}%
\setlength{\unitlength}{4144sp}%
\begingroup\makeatletter\ifx\SetFigFont\undefined%
\gdef\SetFigFont#1#2#3#4#5{%
  \reset@font\fontsize{#1}{#2pt}%
  \fontfamily{#3}\fontseries{#4}\fontshape{#5}%
  \selectfont}%
\fi\endgroup%
\begin{picture}(1349,1688)(2480,-1296)
\put(3279,-1236){\makebox(0,0)[b]{\smash{{\SetFigFont{12}{14.4}{\familydefault}{\mddefault}{\updefault}{\color[rgb]{0,0,0}$\tau$}%
}}}}
\put(2780, -5){\makebox(0,0)[b]{\smash{{\SetFigFont{12}{14.4}{\familydefault}{\mddefault}{\updefault}{\color[rgb]{0,0,0}$c$}%
}}}}
\put(2521,-769){\makebox(0,0)[b]{\smash{{\SetFigFont{12}{14.4}{\familydefault}{\mddefault}{\updefault}{\color[rgb]{0,0,0}$\tau$}%
}}}}
\put(3116,-670){\makebox(0,0)[b]{\smash{{\SetFigFont{12}{14.4}{\familydefault}{\mddefault}{\updefault}{\color[rgb]{0,0,0}$r_0$}%
}}}}
\put(3434,-673){\makebox(0,0)[b]{\smash{{\SetFigFont{12}{14.4}{\familydefault}{\mddefault}{\updefault}{\color[rgb]{0,0,0}$r_1$}%
}}}}
\put(3756,-107){\makebox(0,0)[b]{\smash{{\SetFigFont{12}{14.4}{\familydefault}{\mddefault}{\updefault}{\color[rgb]{0,0,0}$e$}%
}}}}
\end{picture}%